\documentclass[aps,prd,preprint,showpacs]{revtex4}

\usepackage{epsfig}
\usepackage{amssymb}
\usepackage{graphicx}

\newcommand{\Pomeron}{I\!\!P}
\newcommand{\Reggeon}{I\!\!R}

\begin{document}

\title{
Leading twist nuclear shadowing: uncertainties, 
comparison to experiments and higher twist effects}

\author{L. Frankfurt}
\affiliation{Nuclear Physics Dept., School of Physics and Astronomy, Tel Aviv
University, 69978 Tel Aviv, Israel}
\email[]{frankfur@lev.tau.ac.il}
\author{V. Guzey}
\affiliation{Institut f{\"u}r Theoretische Physik II, Ruhr-Universit{\"a}t
 Bochum, D-44780 Bochum, Germany}
\email[]{vadim.guzey@tp2.ruhr-uni-bochum.de}
\author{M. Strikman}
\affiliation{Department of Physics, the Pennsylvania State University, State
  College, PA 16802, USA}
\email[]{strikman@phys.psu.edu}

\begin{abstract}
Using the leading twist approach to nuclear shadowing, which is based on the 
relationship between nuclear shadowing and diffraction on a nucleon,
we calculate next-to-leading order nuclear parton distribution
functions (nPDFs) and structure functions
 in the region $0.2 > x > 10^{-5}$ and $Q^2 \geq 4$ GeV$^2$.
 The uncertainties of our predictions due the uncertainties of the
 experimental input and  the theory are quantified.
We determine the relative role of the small ($\sim Q^2$) and large ($\gg Q^2$)
 diffractive masses in nuclear shadowing as a function of $x$ 
and find that the large mass contribution,
which is an analog of the triple Pomeron exchange, becomes significant
only for $x \le 10^{-4}$.
Comparing our predictions to the available fixed-target nuclear DIS
data, we argue, based on the current experimental studies of the leading 
twist diffraction, 
that  the data at 
moderately 
small $x\sim 0.01$ and $Q^2 \sim 2$ GeV$^2$ 
could
 contain significant
 higher twist
effects hindering the extraction of nPDFs from that data.
Also, we find that the next-to-leading order effects in nuclear
shadowing 
in the ratio of the nucleus to nucleon 
structure functions $F_2$ are quite sizable.
Within the same formalism, we also 
present results for 
 the impact parameter dependence
 of nPDFs.
We also address the problem of extracting of the neutron 
 $F_{2n}(x,Q^2)$ from the deuteron
and proton data. We suggest a 
 simple and nearly model-independent procedure of 
 correcting for nuclear shadowing effects using
$F_2^A/F_2^D$ ratios.
 
 \end{abstract}


\maketitle

\section{Introduction}
\label{sec:intro}

One way to analyze the microscopic structure of atomic nuclei is to study
 the distribution of quarks and gluons,
 as well as their correlations, in nuclei. These nuclear parton distribution
 functions (nPDFs) can be accessed 
using various deep inelastic scattering (DIS) processes: Inclusive scattering
 of leptons, high-mass dimuon 
production using proton beams, exclusive electroproduction of vector mesons.
 None of the above processes
 determines nPDFs comprehensively, only taken together do these experiments
 provide stringent constraints on nPDFs.

The discussion of the present paper is centered around the nuclear effects of
 nuclear shadowing and antishadowing (enhancement),
 which affect nPDFs at small values of Bjorken variable $x$, 
$10^{-5} \leq x \leq 0.2$. 
Nuclear shadowing of nPDFs is developing into  an increasingly important
 subject because it is
 involved in the interpretation of the RHIC data on jet production,
 evaluation of hard phenomena in proton-nucleus
 and nucleus-nucleus collisions at the LHC, estimates of the black 
limit scattering regime in DIS, etc.

The major obstacle that hinders our deeper knowledge of nPDFs at small $x$
is that, up to the present day, all experiments aiming to study nPDFs
are performed with fixed (stationary) nuclear targets. 
In these data, the values of $x$ and $Q^2$ are strongly correlated
and one measures nPDFs essentially
along a curve in the $x-Q^2$ plane rather than exploring the entire plane.
Moreover, for $Q^2 > 1$ GeV$^2$, the data cover the 
region $x > 5 \times 10^{-3}$, where the effect of nuclear
 shadowing is just setting in.
 As a result, when one attempts to globally fit the available data
 by modeling nPDFs
 at some initial scale $Q_0^2$ 
and then performing QCD evolution, various groups 
\cite{Eskola,Kumano,HIJING,Eskola:last,deFlorian} produce 
significantly different results.

An alternative to the fitting to the data is to combine
 the Gribov theory \cite{Gribov}, 
which relates the nuclear shadowing correction to the total hadron-deuteron
 cross section to the cross section of
diffraction off a free nucleon, 
with
 the Collins factorization theorem~\cite{Fact:diffractive}
 for hard
 diffraction in DIS.
 The resulting leading twist theory 
 of nuclear shadowing was developed in \cite{FS99} and later
 elaborated on in \cite{FGMS02}.

The Gribov theory has been applied to the description of nuclear shadowing for
many years. First it was done in the region of small $Q^2$, where generalized 
vector dominance
model gives a good description of diffraction,  see review in
\cite{BPY}, and later in deep inelastic region, where large diffractive
masses $M^2 \propto Q^2$ dominate \cite{FS:PRep160}.
 A number of successful model
calculations were performed 
\cite{Thomas,Ratzka,Piller:review,Badelek,Shaw,FS89,Nikolaev,Povh,Lu}
before the experimental data from HERA
became available. A calculation constrained to reproduce the HERA data
using the Gribov theory was presented
in \cite{Kaidalov}. It focuses on the calculation of
nuclear shadowing for $F_{2}^A$ at intermediate $Q^2$ where 
leading and higher twist effects are equally important. A
fair agreement of the data with the Gribov theory has been found.
 However, this approach
does not 
involve the use of the Collins factorization theorem and, 
hence, does not address 
nPDFs (see a detailed comparison in Sect.~\ref{sec:results}).

The present work extends the calculation of nPDFs of \cite{FGMS02} with
 an emphasis on 
 the theoretical ambiguity
 and accuracy of the predictions and makes a comparison to fixed-target nuclear DIS data.
 In particular, we demonstrate that
\begin{itemize}
{\item The theory of leading twist nuclear shadowing and QCD analysis of hard 
diffraction at HERA
enable one to predict in a model-independent
 way the next-to-leading order nPDFs for $10^{-5} \leq x \lesssim 10^{-2}$ with 30\% 
accuracy, 
Fig.~\ref{fig:input}.
For larger $x$, $10^{-2} \leq x \leq 0.1-0.2$, 
there appears an additional effect of nuclear antishadowing
 that requires modeling 
and whose uncertainty is larger.
In addition, the HERA diffractive data for $x_{\Pomeron} > 0.01$ contains a
sub-leading Reggeon contribution, which adds additional ambiguity to our 
predictions,
especially
 for  $x > 0.01$.
}

{\item The interactions with $N \ge 3$  nucleons (which is a model-dependent
 element of the Gribov approach) give negligible contribution in the NMC
 fixed-target nuclear
DIS kinematics, see Fig.~\ref{fig:rescattering}.  The $A$-dependence of
 the NMC data for $x \sim 0.01$ is reasonably well
reproduced, see Fig.~\ref{fig:adep}. 
}
{\item The failure to describe the
absolute value of the 
$F_2^A/(A F_2^N)$
ratios of the
 available 
fixed-target data for $0.003 < x <0.02$ and $Q^2 < 3$ GeV$^2$
likely
 indicates the presence of significant higher twist effects in the data.
Indeed,  when
the leading twist shadowing is complemented by
 higher twist effects, which are modeled by $\rho$, $\phi$ and $\omega$ meson
contributions in the spirit of vector meson dominance,
the agreement with the data becomes fairly good, 
see Figs.~\ref{fig:nmcc12}, \ref{fig:nmcca40} and \ref{fig:nmcpb}.
All this
 signals that any leading twist QCD analysis of the available data 
is unreliable
 for $0.003 < x <0.02$.}
{\item The next-to-leading order (NLO) effects in the 
 $F_2^A/(A F_2^N)$ ratios
are found to be quite sizable.
This means that it is not self-consistent to use the leading order
 parameterizations of nPDFs 
in the NLO QCD calculations, see Fig.~\ref{fig:f2eskola}. 
}
\end{itemize}

In short, the main goals of the paper are to give a concise summary of the
 leading twist theory of nuclear shadowing,
to assess the theoretical uncertainties of the resulting predictions and
 to make a comparison to nuclear DIS data.
We attempted to give a self-contained presentation and, hence,
 this paper
can be rightfully considered as a {\it guide}
to leading twist nuclear shadowing.

 \section{Leading twist theory of nuclear shadowing}
\label{sec:theory}

In this section, we review the leading twist approach to nuclear shadowing
 developed in \cite{FS99} and
further elaborated on in  \cite{FGMS02}.
\begin{figure}
\includegraphics[width=12cm,height=12cm]{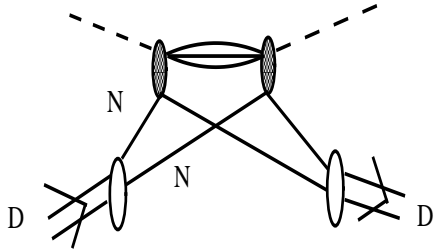}
\vskip -3cm
\caption{Gribov's theorem \protect\cite{Gribov}: The forward hadron-deuteron rescattering amplitude,
 which gives rise to nuclear shadowing, is proportional to the differential hadron-nucleon diffractive cross section
 at $t\sim 0$.
 }
\label{fig:gribov}
\end{figure}

The approach is based on the 1969 work by  V.~Gribov \cite{Gribov}, where the following theorem was proven.
 Let us consider hadron-deuteron scattering at high energies within the
 approximation that the radius of
 the deuteron is much larger than the range of the strong interaction.
Then the shadowing correction to the total cross section is expressed in
 terms of the differential
 diffractive hadron-nucleon cross section. This is demonstrated in
 Fig.~\ref{fig:gribov}:
 The forward hadron-deuteron rescattering amplitude giving rise to the
 nuclear shadowing correction
 contains the hadron-nucleon diffractive amplitude (denoted by the shaded blob) squared.

The relationship between nuclear shadowing and 
diffraction was used in the analysis of 
parton densities of 
deuterium and
 other nuclei by Frankfurt and Strikman in \cite{FS99}.
 For deuterium and other sufficiently light (low nuclear density) nuclei,
 nuclear  shadowing and diffraction on the nucleon are related in a
 model-independent way using the Gribov theorem~\footnote{It is possible to derive  the Gribov theorem including
  corrections due to the real part of the diffractive amplitude (those
  were neglected in the Gribov's original approach based on the
  Pomeron model with $\alpha_{\Pomeron}(0)=1$) using the Abramovskii,
  Gribov and Kancheli (AGK) cutting rules \cite{AGK}, see
  \cite{FS99,FGMS02}. Hence, 
in the small-$x$ limit ($x \ll 10^{-2}$),
 the relation between 
  shadowing and diffraction is essentially a consequence of unitarity.}.

The
 generalization to heavy nuclei involves certain 
modeling of multiple rescattering contributions,
 which, however, is
 under control~\cite{Alvero}. Below we shall recapitulate 
the derivation of the leading twist nuclear shadowing for nPDFs, which can
 be carried out in three steps. 

\underline{Step 1}. The shadowing correction arising from the coherent interaction with
 any two nucleons of the nuclear target with the atomic mass number $A$, 
$\delta F_{2A}^{(2)}$
 (the superscript $(2)$ serves as a reminder that only the interaction with two nucleons
 is accounted for),
 is expressed in terms of the proton  diffractive structure function
 $F_2^{D(4)}$
 (the superscript $(4)$ indicates the dependence on four kinematic variables)
as a result of the generalization of the Gribov result for deuterium (see
also Ref.~\cite{BPY}). This
does not require decomposition over  twists and is
therefore valid even  for the case of real photon interactions. The shadowing
 correction $\delta F_{2A}^{(2)}$ 
reads~\footnote{The real part of the amplitude was neglected in  \cite{Kaidalov}, and,
 therefore,
 their corresponding expressions  do not contain the factor 
$(1-i \eta)^2/(1+\eta^2)$.}
\begin{eqnarray}
&&\delta F_{2A}^{(2)}(x,Q^2)=\frac{A(A-1)}{2} 16 \pi {\cal R}e \Bigg[\frac{(1-i\eta)^2}{1+\eta^2} \int d^2 b \int^{\infty}_{-\infty} dz_1 \int^{\infty}_{z_1} dz_2 \int^{x_{\Pomeron,0}}_{x} d x_{\Pomeron} \nonumber\\
&&\times F_2^{D(4)}(\beta,Q^2,x_{\Pomeron},t)\big|_{t=t_{{\rm min}}} \rho_A(b,z_1) \rho_A(b,z_2) e^{i x_{\Pomeron} m_N (z_1-z_2)} \Bigg] \,,
\label{eq:step1} 
\end{eqnarray}
with $\eta$ the ratio of the real to imaginary parts of the diffractive scattering amplitude;
 $z_1$, $z_2$ and $\vec{b}$ the longitudinal (in the direction of the incoming virtual photon)
 and transverse coordinates of the nucleons involved (defined with respect to the nuclear center);     
 $\beta$, $x_{\Pomeron}$ and $t$ the usual kinematic variables used in diffraction.
 Throughout this work, we use $\beta=x/x_{\Pomeron}$.
Equation~(\ref{eq:step1}) uses the fact that the $t$-dependence of the elementary diffractive
 amplitude is much weaker than that of the nuclear wave function, and, hence, $F_2^D(4)$ 
can be approximately evaluated at $t=t_{{\rm min}} \approx 0$. All information about the
 nucleus is encoded in the nucleon distributions $\rho_A(b,z_i)$, see
Appendix~\ref{sec:appendix} for details. Finally, $x_{\Pomeron,0}$ is
 a cut-off parameter ($x_{\Pomeron,0}=0.1$ for quarks and $x_{\Pomeron,0}=0.03$ for gluons),
 which will be discussed later in the text.

\begin{figure}
\includegraphics[width=12cm,height=12cm]{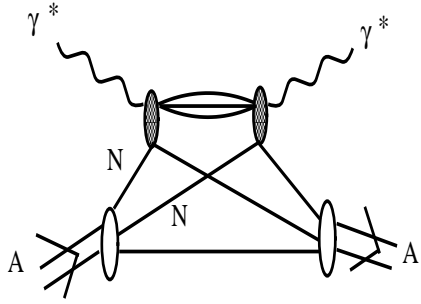}
\vskip -3cm
\caption{The forward $\gamma^{\ast}$-nucleus rescattering amplitude that gives
 the principal contribution to nuclear shadowing.}
\label{fig:gribov2}
\end{figure}

 The origin of all factors in Eq.~(\ref{eq:step1}) can be readily seen by
 considering the corresponding
 forward double rescattering Feynman diagram (see Fig.~\ref{fig:gribov2}),  
which accounts for the diffractive production of intermediate hadronic states by the incoming virtual photon:
\begin{itemize}
{\item The combinatoric factor $A(A-1)/2$ is the number of the pairs of nucleons involved in the rescattering process.}
{\item  The factor $16 \pi$ provides the correct translation of the differential diffractive to 
the total rescattering cross section (see the definition later), as required by the Glauber theory \cite{BPY,Glauber}.}
{\item The factor $(1-i \eta)^2/(1+\eta^2)$ is a correction for the real part of the diffractive
 scattering amplitude ${\cal A}$. Since the shadowing correction is proportional to $(Im{\cal A})^2$,
 while the total diffractive cross section is proportional to $|{\cal A}|^2$,
 the factor $(1-i \eta)^2/(1+\eta^2)$ emerges naturally, when one expresses nuclear shadowing in terms
 of the total diffractive cross section (diffractive structure function).}
{\item The integration over the positions of the nucleons is the same as in the Glauber theory.
Similarly,
because the recoil of the nucleons is neglected (the transverse 
radius of the elementary strong amplitude is much smaller than the  scale of the variation of the nuclear density),
 both involved nucleons have the same 
transverse coordinate $\vec{b}$.}
{\item The integration over $x_{\Pomeron}$ represents the sum over the masses of the
 diffractively produced intermediate states.}
{\item In order to contribute to nuclear shadowing
 (not to break up the nucleus
in its transition from the $|{\rm in} \rangle$-state  to the $\langle {\rm out}|$-state),
 the virtual photon 
should interact with the nucleons diffractively. The product of the two diffractive amplitudes
 (depicted as shaded blobs in Fig.~\ref{fig:gribov2}) gives the diffractive structure function
 of the proton $F_2^{D(4)}$. Also note that we do not distinguish between diffraction on the proton
 and neutron in the present work,
as the corresponding diffractive amplitudes are equal at small $x$.
}
{\item The effect of the nucleus is given by the nucleon densities $\rho_A(b,z_i)$.
 For the sufficiently heavy nuclei that we consider, nucleon-nucleon correlations can be neglected
 and the nuclear wave function squared can be approximated well by the  product of individual
 $\rho_A(b,z_i)$ for each nucleon (the so-called independent particle approximation).} 
{\item The factor $e^{i x_{\Pomeron} m_N (z_1-z_2)}$ is a consequence of the propagation
 of the
 diffractively produced intermediate state between the two nucleons involved.}
\end{itemize}

\underline{Step 2}. The QCD factorization theorems for inclusive~\cite{Fact:inclusive}
 and hard diffractive  
DIS~\cite{Fact:diffractive} can be used to relate the structure functions in Eq.~(\ref{eq:step1}) 
to the corresponding -- inclusive and diffractive -- parton distribution functions.
 Since the coefficient functions (hard scattering parts) are the same for both inclusive and diffractive
 structure functions, the relation between the shadowing correction to nPDFs and the proton diffractive
 parton distribution functions (PDFs) is  given by an equation similar to Eq.~(\ref{eq:step1}).
 The shadowing correction to the nPDF of flavor $j$, $f_{j/A}$, $\delta f_{j/A}^{(2)}$, is related to the
 proton (nucleon) diffractive PDF $f_{j/N}^{D(4)}$ of the same flavor
\begin{eqnarray}
&&\delta f_{j/A}^{(2)}(x,Q^2)=\frac{A(A-1)}{2} 16 \pi {\cal R}e \Bigg[\frac{(1-i\eta)^2}{1+\eta^2} \int d^2 b \int^{\infty}_{-\infty} dz_1 \int^{\infty}_{z_1} dz_2 \int^{x_{\Pomeron,0}}_{x} d x_{\Pomeron} \nonumber\\
&&\times f_{j/N}^{D(4)}(\beta,Q^2,x_{\Pomeron},t)\big|_{t=t_{{\rm min}}} \rho_A(b,z_1) \rho_A(b,z_2) e^{i x_{\Pomeron} m_N (z_1-z_2)} \Bigg] \,.
\label{eq:step2} 
\end{eqnarray}
Equation~(\ref{eq:step2}) is very essential in several ways. Firstly, it enables one to evaluate  nuclear
 shadowing 
for each parton flavor $j$ separately. Secondly, since the diffractive PDFs obey leading twist QCD evolution,
 so does the shadowing correction  $\delta f_{j/A}^{(2)}$. 
This explains why the
 considered theory can
 be legitimately called the leading twist approach.
Since Eq.~(\ref{eq:step2}) is based on the QCD factorization theorem,
it is valid to all orders in $\alpha_s$. Hence, if $f_{j/N}^{D(4)}$ is
known with the next-to-leading order (NLO) accuracy, as is the case
for the used H1 parameterization for $f_{j/N}^{D(4)}$, we can readily
make predictions for NLO nPDFs.

\begin{figure}
\includegraphics[width=12cm,height=12cm]{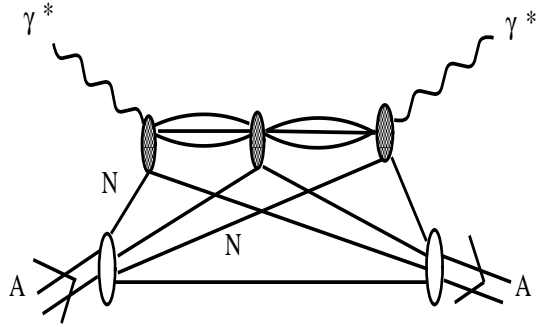}
\vspace{-3cm}
\caption{The forward $\gamma^{\ast}$-nucleus triple scattering amplitude.}
\label{fig:gribov3}
\end{figure}

\underline{Step 3}. Equation~(\ref{eq:step2}) is derived in the approximation
 of the low nuclear thickness,
 and it takes into account only the interaction with two nucleons of the target. The effect of the rescattering
 on three and more nucleons can be taken into account by introducing  the attenuation 
factor $T(b,z_1,z_2)$ (see for example \cite{BPY}),
\begin{equation}
T(b,z_1,z_2)=e^{-(A/2)(1-i\eta)\sigma_{{\rm eff}}^j \int_{z_1}^{z_2} dz \rho_A(b,z)} \,,
\label{eq:step3} 
\end{equation}
where the meaning of $\sigma_{{\rm eff}}^j$ should become clear after the following discussion.
 Let us consider sufficiently small values of Bjorken variable $x$ such that the factor
 $e^{i x_{\Pomeron} m_N (z_1-z_2)}$ in Eq.~(\ref{eq:step2}) can be neglected.
 Then, introducing $\sigma_{{\rm eff}}^j$ as
\begin{equation}
\sigma_{{\rm eff}}^j(x,Q^2)=\frac{16 \pi}{f_{j/N}(x,Q^2)(1+\eta^2)}\int_x^{x_{\Pomeron,0}}d x_{\Pomeron}  f_{j/N}^{D(4)}(\beta,Q^2,x_{\Pomeron},t)\big|_{t=t_{{\rm min}}} \,,
\label{eq:sigma} 
\end{equation}
Eq.~(\ref{eq:step2}) can be written in the form equivalent to the usual Glauber approximation
\begin{eqnarray}
\delta f_{j/A}^{(2)}(x,Q^2) & \approx & \frac{A(A-1)}{2} (1-\eta^2)\sigma_{{\rm eff}}^j(x,Q^2) f_{j/N}(x,Q^2) \nonumber\\
&&\times  \int d^2 b \int^{\infty}_{-\infty} dz_1 \int^{\infty}_{z_1} dz_2   \rho_A(b,z_1) \rho_A(b,z_2)  \,,
\label{eq:step2appr} 
\end{eqnarray}
 where $f_{j/N}$ is the proton inclusive PDF. Therefore, it is clear that thus introduced
 $\sigma_{{\rm eff}}^j$ has the meaning of the rescattering cross section, which determines
 the amount of nuclear shadowing in the approximation of Eq.~(\ref{eq:step2appr}). 
Hence, it is natural to assume that the same cross section describes rescattering with the
 interaction with three and more nucleons, as postulated by the definition of the attenuation
 factor  $T(b,z_1,z_2)$ by Eq.~(\ref{eq:step3}). In the language of 
Feynman diagrams, the assumed form
 of the attenuation factor implies that the diffractively produced
 intermediate state rescatters 
without a significant change of mass
with the same cross section on all remaining nucleons of the target,
 as depicted in Fig.~\ref{fig:gribov3}
 for the case of the triple scattering. 
The approximation of elastic rescattering is not important
at small enough $x$ where longitudinal distances are much larger than 
the nuclear
size, see discussion below.

Corrections to the elastic rescattering approximation can be estimated 
by taking into account
the effects of fluctuations of  the strength of the rescattering interaction.
Modeling of these effects was performed in \cite{Alvero} with the
 conclusion that for  
a wide range of cross section fluctuations, the reduction of  nuclear shadowing
(for fixed $\sigma_{{\rm eff}}$ ) remains a rather small correction for all nuclei.

After introducing the attenuation factor into Eq.~(\ref{eq:step2}), the complete expression for the shadowing 
correction, $\delta f_{j/A}$, becomes
\begin{eqnarray}
&&\delta f_{j/A}(x,Q^2)=\frac{A(A-1)}{2} 16 \pi {\cal R}e \Bigg[\frac{(1-i\eta)^2}{1+\eta^2} \int d^2 b \int^{\infty}_{-\infty} dz_1 \int^{\infty}_{z_1} dz_2 \int^{x_{\Pomeron,0}}_{x} d x_{\Pomeron}  \nonumber\\
&&\times f_{j/N}^{D(4)}(\beta,Q^2,x_{\Pomeron},t_{{\rm min}}) \rho_A(b,z_1) \rho_A(b,z_2) e^{i x_{\Pomeron} m_N (z_1-z_2)} e^{-(A/2)(1-i\eta)\sigma_{{\rm eff}}^j \int_{z_1}^{z_2} dz \rho_A(b,z)} \Bigg] \,.
\label{eq:master} 
\end{eqnarray}

This is our master equation (see also Eq.~(\ref{eq:master2})).
It contains several sources of model-dependence and theoretical ambiguity.
 First, 
the attenuation factor $T(b,z_1,z_2)$
assumes that multiple rescatterings can be described by a single rescattering
 cross section~\footnote{
The double rescattering term can be calculated using Eq.~(\ref{eq:step2})
 at any $Q^2$. However, the quasieikonal approximation employed in
 Eq.~(\ref{eq:master}) is  best justified at low $Q^2 \sim Q_0^2$, where
 fluctuations in the strength of the interaction are smaller (see the discussion in
 Ref.~\cite{FGMS02}).
 The
 QCD evolution equations automatically account for the proper increase of the
 fluctuations of the effective cross section around its average value  $\sigma_{{\rm eff}}^j$
 with an increase of $Q^2$. This important effect is omitted if
one attempts to
 apply Eq.~(\ref{eq:master}) at  $Q^2 >Q_0^2$ with a $Q^2$-dependent $\sigma_{eff}^j(Q^2)$.}
 $\sigma_{\rm eff}^j$, i.e.
cross section fluctuations are neglected in the interaction with three and more nucleons.
Note that in the phenomenologically important kinematic region of fixed-target 
experiments, $x > 0.01$ and $Q^2 > 2$ GeV$^2$, the uncertainty associated with
 the attenuation factor $T(b,z_1,z_2)$ is negligible since the rescattering contribution 
to shadowing  is small, see Fig.~\ref{fig:rescattering}.
Second, the necessity to introduce the parameter $x_{\Pomeron,0}$
 is a consequence of the fact
 that Eq.~(\ref{eq:master})
applies only to the region of nuclear shadowing: The transition to the region of the enhancement of
 nPDFs should be modeled separately. This is the role of the parameter $x_{\Pomeron,0}$.
Third, there are experimental uncertainties in the determination of the
 diffractive PDFs
$f_{j/N}^{D(4)}$ which we use as an input in Eq.~(\ref{eq:master}).

Equation~(\ref{eq:master}) defines the input nPDFs for the DGLAP 
evolution equations. 
As a  starting evolution scale 
$Q_0^2$, we take $Q_0^2=4$ GeV$^2$: This is the lowest value of $Q^2$ of the H1 diffractive 
fit~\cite{H1:parametrization}. Nuclear PDFs at $Q^2 > Q_0^2$ are obtained using the NLO QCD
 evolution equations. Therefore, we predict that nuclear shadowing is a leading twist phenomenon.

In the small-$x$ limit, which for practical purposes means $x < 10^{-3}$,
 the factor 
$e^{i x_{\Pomeron} m_N (z_1-z_2)}$ in Eq.~(\ref{eq:master}) can be safely 
omitted, which results in a
significant simplification of the master formula (after integration by parts two times)
\begin{equation}
\delta f_{j/A}(x,Q^2)=\frac{2 \left(1-1/A \right)f_{j/N}(x,Q^2)}{\sigma_{{\rm eff}}^j} Re 
\left(\int d^2 b \left(e^{-L\, T(b)}-1+L\, T(b) \right) \right) \,,
\label{eq:gl}
\end{equation}
where $L=A/2 \, (1-i\eta) \, \sigma_{{\rm eff}}^j$; $T(b)=\int^{\infty}_{-\infty} dz \,\rho_A(b,z)$.

In the heavy nucleus limit ($ A \to \infty$) and at fixed $\sigma_{{\rm eff}}$,
\begin{equation}
\frac{f_{j/A}(x,Q^2)}{A f_{j/N}(x,Q^2)}=1- \frac{\delta f_{j/A}(x,Q^2)}{A f_{j/N}(x,Q^2)}=
\frac{2\pi R_A^2}{A \sigma_{{\rm eff}}^j} \,,
\end{equation}
where $R_A$ is the nuclear size. As can be seen, in the  theoretical limit of infinitely 
heavy nucleus, nuclear shadowing equals the ratios of the nuclear to nucleon sizes, i.e.
it is a purely geometrical effect. 
At the  same time, in the $Q^2=const,\, W\to \infty$ limit,  when the 
leading twist approximation is 
violated and the radius of the strong interaction becomes larger than $R_A$,
 the $\sigma_{\gamma^{\ast} A}/\sigma_{\gamma^{\ast} N}$ ratio should approach unity \cite{FSZ04}.

\section{Parameters and uncertainties of the method}
\label{sec:using}

The master equation~(\ref{eq:master}) uses as  input
 the information on hard diffraction in DIS on the proton, which was measured at HERA by ZEUS~\cite{ZEUS:1994} 
and H1~\cite{H1:1994} collaborations. We use the H1 parameterization of  
$f_{j/N}^{D(3)}$~\cite{H1:parametrization}
 (note the superscript $(3)$ indicating that the $t$-dependence of diffraction is not measured),
which is based on the QCD analysis of the 1994 H1 data \cite{H1:1994} (we use 
Fit B, 
see Appendix A of Ref.~\cite{FGMS02}). The choice of the H1 parameterization is motivated by the
 following observations:
\begin{itemize}
{\item It is available in an easily accessible and usable form, see \cite{H1:parametrization} and 
also Appendix A of Ref.~\cite{FGMS02}.}
{\item  The diffractive jet production in DIS at HERA data \cite{H1:jets} is best described by 
the H1 parameterization.
 The fit of Alvero, Collins, Terron and Whitmore \cite{ACTW} somewhat overestimates the data.
 Another parameterization available in the literature, that of Hautmann, Kunszt and Soper \cite{HKS}, 
is not based on the detailed fit to the available diffractive data.}
{\item The 1994 H1 fit is in a fair agreement with the most recent 1997 H1 data \cite{H1:1997}.
However, the 1997 H1 data indicates that the gluon distribution of the 1994 fit is too large by 
about 25\%. Hence, in our analysis we multiplied the gluon diffractive distribution of 
 \cite{H1:parametrization} by 0.75. 
}
\end{itemize}

Since the diffractive PDF $f_{j/N}^{D(4)}$ enters Eq.~(\ref{eq:master}) at $t \approx 0$, 
one has to assume a certain $t$-dependence in order to be able to use the H1 results for the $t$-integrated
  $f_{j/N}^{D(3)}$. The common choice is to
 assume that
\begin{equation}
f_{j/N}^{D(4)}(\beta,Q^2,x_{\Pomeron},t)=e^{B_j t} f_{j/N}^{D(4)}(\beta,Q^2,x_{\Pomeron},t \approx 0) \,,
\end{equation}
so that after the integration over $t$, one obtains 
\begin{equation}
f_{j/N}^{D(4)}(\beta,Q^2,x_{\Pomeron},t \approx 0)=B_j f_{j/N}^{D(3)}(\beta,Q^2,x_{\Pomeron}) \,,
\label{eq:43} 
\end{equation}
where $B_j$ is the slope of the $t$-dependence of $f_{j/N}^{D(4)}$.
A priory there is no reason why the slope $B_j$ should be equal for all parton flavors $j$ and, 
hence, we introduce its explicit flavor dependence.
In our analysis we use the following values for  $B_j$. For all quark flavors,
 we use 
$B_q=7.2 \pm 1.1({\rm stat.})^{+0.7}_{-0.9}({\rm syst.})$ GeV$^{-2}$,
 which is determined by
 the measurement of the $t$-dependence of the diffractive structure function 
$F_2^{D(4)}$,
 as measured by the ZEUS collaboration \cite{ZEUS:tslope}.
Of course, the diffractive slope should increase with decreasing $x$
(diffractive cone shrinkage). However, since the experimental error of
the value of $B_q$ is large
and no measurements of the $x$-dependence of $B_q$ are available, 
 any theoretically expected logarithmic increase of $B_q$ will be within
 the quoted experimental
 errors. Hence, it is sufficient to use the $x$-independent $B_q$.

The slope of the gluon PDF, $B_g$, could be different from $B_q$.
If the gluon-induced diffraction is dominated by small-size (compared to
typical soft physics sizes) partonic configurations in the
projectile, $B_g$ could as low 
as the slope of  $J/\psi$ diffractive production measured at HERA which 
is substantially  lower
than $B_q$. To reflect the uncertainties in the value of $B_g$ we
examined two scenarios: $B_g=4+0.2 \ln(10^{-3}/x)$ GeV$^{-2}$ and
$B_g=6+0.25 \ln(10^{-3}/x)$ GeV$^{-2}$. 
The first one corresponds to the lower end of the values of the $J/\psi$ 
photoproduction slope
reported at HERA \cite{ph1}, while the second one is close to $B_q$ and 
to the $J/ \psi$ slope reported in \cite{ph2}.

The 
analysis of 
 recent ZEUS data on the slope of $F_2^{D(4)}$ \cite{ZEUS2004} reports
 the values similar to
those reported in \cite{ZEUS:tslope}. 
The analysis
 also provides information on the $x_{\Pomeron}$-dependence of the slope
 though not on the $\beta$-dependence of the slope (average  $\beta$ for 
the data sample is growing with a decrease of 
$x_{\Pomeron}$). Overall, the data appear to be consistent with the Regge 
factorization and for the most of the $x_{\Pomeron}$ range,
the  gluon diffractive PDF appears to give a significant, if not the dominant,
 contribution. This suggests that our model with a higher value of $B_g$
 is closer to the data, though in view of the lack of the data on the 
slope of the gluon induced diffraction at large $\beta$, we feel necessary
 to  keep the lower $B_g$ model as well.

The analysis of the 1994 H1 data \cite{H1:1994} showed that at large
 $x_{\Pomeron}$, the successful fit
to the data requires both the Pomeron and the Reggeon contributions.
 In our numerical 
analysis we include only the 
dominant Pomeron part, because
the subleading Reggeon contribution begins to play a role only for $x > 0.01$,
where the theoretical ambiguities are large anyway. Therefore,
in our analysis we use the Reggeon contribution only to estimate its 
contribution to the overall uncertainty of our predictions, 
see Appendix~\ref{sec:reggeon} for details.

Using Eq.~(\ref{eq:43}), 
 the rescattering cross section $\sigma_{{\rm eff}}^j$ becomes
\begin{equation}
\sigma_{{\rm eff}}^j(x,Q^2)=\frac{16 \pi B_j}{f_{j/N}(x,Q^2)(1+\eta^2)}
\int_x^{x_{\Pomeron,0}}
d x_{\Pomeron}  f_{j/ N}^{D(3)}(\beta,Q^2,x_{\Pomeron})   \,,
\label{eq:sigma2} 
\end{equation}
where $\eta$ is the ratio of the real to imaginary parts of the diffractive amplitude ${\cal A}$.
 This ratio can be related to the intercept of the effective Pomeron trajectory, $\alpha_{\Pomeron}(0)$,
  using the Gribov-Migdal result \cite{Migdal}
\begin {equation}
\eta \approx \frac{\pi}{2}\big(\alpha_{\Pomeron}(0)-1\big)=0.32 \,,
\end{equation}
where the H1 value for $\alpha_{\Pomeron}(0)$ was employed.

\begin{figure}
\includegraphics[width=13cm,height=13cm]{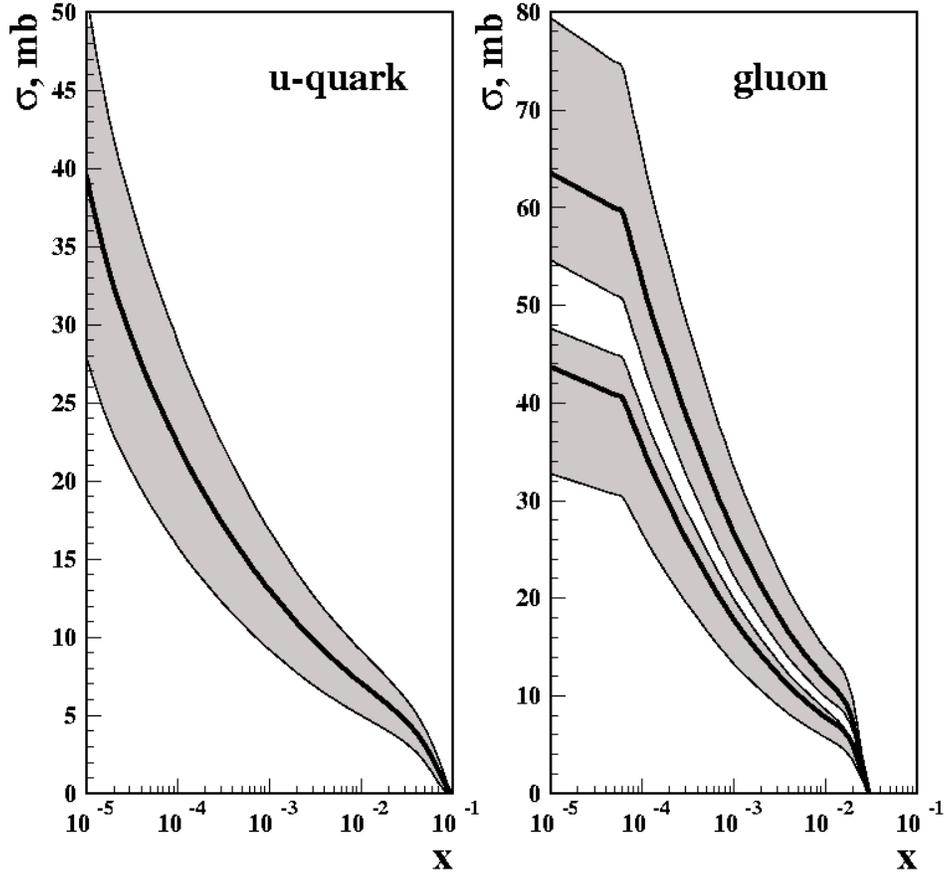}
\caption{The effective cross section $\sigma_{{\rm eff}}$ for the
  anti $u$-quark and gluon channels
at $Q_0^2=4$ GeV$^2$.
The error bands represent the uncertainty in the predictions discussed in the text.}
\label{fig:sigma}
\end{figure}

The results of the evaluation of $\sigma_{{\rm eff}}^j$
are presented in Fig.~\ref{fig:sigma}. The left panel presents $\sigma_{{\rm eff}}$ for 
anti $u$-quarks and the right panel is for the gluons, both cases  for $Q^2=4$ GeV$^2$. 
The error bands around the central curves represent the uncertainty in the determination of
$\sigma_{{\rm eff}}$. This uncertainty comes from the uncertainties in $B_q$, 
$f_{j/N}^{D(3)}(\beta, Q^2, x_{\Pomeron})$ (taken to be 25\%)
 and the choice of $x_{\Pomeron,0}$ added in quadrature.
Two solid curves for the gluon case correspond to the two scenarios for the slope $B_g$ discussed above.

Now we would like to examine which values of the diffractive masses or 
$\beta$ contribute
to $\sigma_{{\rm eff}}$. 
At very high energies (small $x$), one enters the regime analogous to the
triple Pomeron limit of hadronic physics, which corresponds to $\beta =Q^2/(Q^2+M_X^2) \ll 1$.
In this case, 
one may need to resum logs of energy in the diffractive block 
(logs of $\beta$).
However, deviations from DGLAP are
expected only for $\beta \le 10^{-3}$, which is beyond the $x$ range that
we consider~\cite{Ciafaloni}.
At extremely small $x$, the contribution of small $\beta$ may
become dominant. This contribution was evaluated within 
Color Condensate model in
\cite{McLerran:4lectures} neglecting the large $\beta$ contribution.

\begin{figure}
\includegraphics[width=13cm,height=13cm]{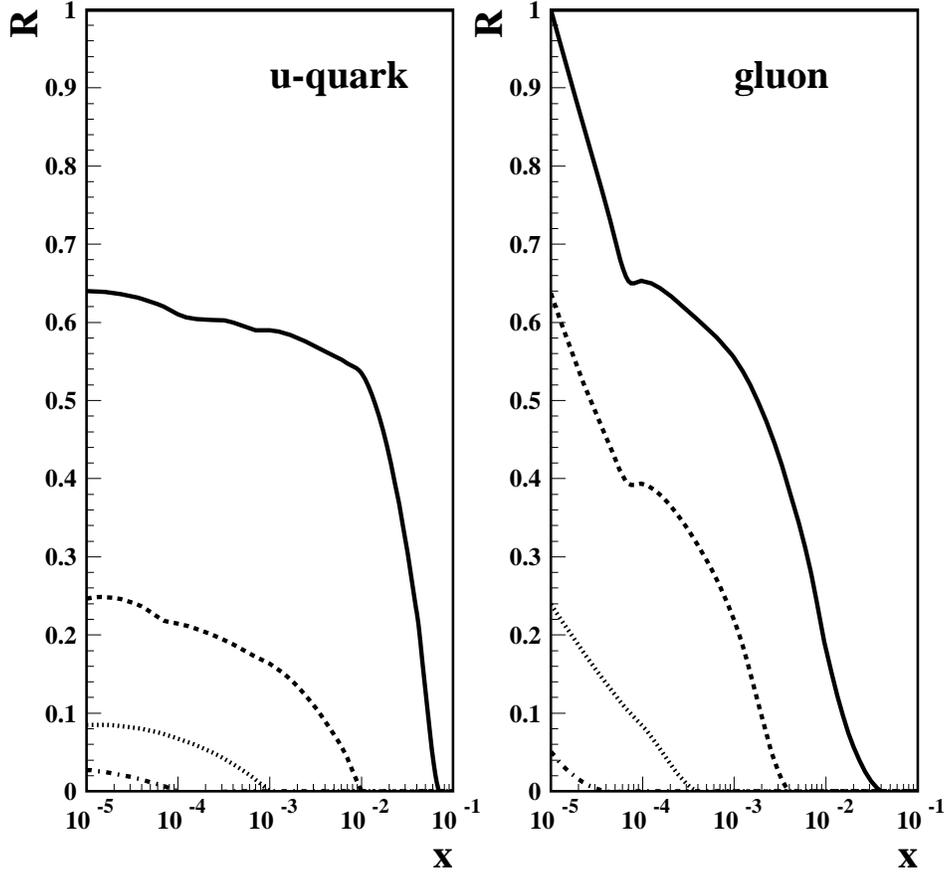}
\caption{The ratio $R$ at $Q_0^2=4$ GeV$^2$. The solid curves correspond to $\beta_{{\rm max}}=0.5$;
the dashed curves correspond to $\beta_{{\rm max}}=0.1$; the dotted curves correspond
 to $\beta_{{\rm max}}=0.01$; the dot-dashed curves correspond to $\beta_{{\rm max}}=0.001$.}
\label{fig:r}

\end{figure}

To analyze at what $x$ small $\beta$ become
dominant, it is convenient to introduce
 the ratio $R$ defined as follows
\begin{equation}
R(\beta_{{\rm max}},x) = \frac{\int_x^{x_{\Pomeron,0}}
d x_{\Pomeron}  f_{j/ N}^{D(3)}(\beta,Q^2,x_{\Pomeron}) \Theta(\beta_{{\rm max}}-\beta)}
{\int_x^{x_{\Pomeron,0}}
d x_{\Pomeron}  f_{j/ N}^{D(3)}(\beta,Q^2,x_{\Pomeron})} \,.
\label{eq:r}
\end{equation}
The ratio $R$ for $u$-quark and gluon channels at $Q_0^2=4$ GeV$^2$ is presented in 
Fig.~\ref{fig:r}. In the figure, the solid curves correspond to $\beta_{{\rm max}}=0.5$;
the dashed curves correspond to $\beta_{{\rm max}}=0.1$; the dotted curves correspond
 to $\beta_{{\rm max}}=0.01$; the dot-dashed curves correspond to $\beta_{{\rm max}}=0.001$.

From Fig.~\ref{fig:r} one can see
 how much  different 
$\beta $-regions
contribute to nuclear shadowing.
 For instance, 
taking $x=10^{-3}$, which roughly corresponds to the 
smallest $x$ which could be reached in the
RHIC kinematics, one sees that 
large diffractive masses that correspond to $\beta \leq 0.1$ (dashed curve) 
contribute 20\% to nuclear 
shadowing in the quark channel and 30\% to nuclear shadowing
 in the gluon channel.
Therefore, Fig.~\ref{fig:r} indicates that if the Color Glass Condensate model is implemented 
in a way consistent with the HERA diffractive data, it  predicts
a very small fraction of total shadowing  for the RHIC kinematic range.

For completeness, we rewrite our master equation, Eq.~(\ref{eq:master}), in the form which
explicitly  includes the diffractive slope $B_j$
\begin{eqnarray}
&&\delta f_{j/A}(x,Q^2)=\frac{A(A-1)}{2} 16 B_j \pi {\cal R}e \Bigg[\frac{(1-i\eta)^2}{1+\eta^2} 
\int d^2 b \int^{\infty}_{-\infty} dz_1 \int^{\infty}_{z_1} dz_2 \int^{x_{\Pomeron,0}}_{x} d x_{\Pomeron}  
\nonumber\\
&&\times f_{j/N}^{D(3)}(\beta,Q^2,x_{\Pomeron}) \rho_A(b,z_1) \rho_A(b,z_2) e^{i x_{\Pomeron} m_N (z_1-z_2)} e^{-(A/2)(1-i\eta)\sigma_{{\rm eff}}^j \int_{z_1}^{z_2} dz \rho_A(b,z)} \Bigg] \,.
\label{eq:master2} 
\end{eqnarray}

We would like to point out that while the leading twist theory of nuclear shadowing is
 applicable
 to the partons of all flavors, see Eq.~(\ref{eq:master2}), using the low-$x$ HERA diffractive data,
 which is heavily dominated by the Pomeron contribution, we cannot make any quantitative
 predictions for nuclear shadowing of the valence quarks in nuclei. Nuclear shadowing for
 the valence quarks is driven by the $t$-channel exchanges with non-vacuum quantum numbers
 (Reggeon contribution), whose contribution is largely lost in the kinematic region of the HERA data.
 In practical terms, this means that   
Eq.~(\ref{eq:master2}) should be applied to evaluate nuclear shadowing for the antiquarks and gluons only.

As mentioned above, Eq.~(\ref{eq:master2}) cannot describe nuclear modifications of PDFs at
 $x > 0.1$ for the quarks and $x > 0.03$ for the gluons, where nuclear antishadowing and the
 EMC effects dominate. For a comprehensive picture of nuclear modification for all values
 of $x$, 
we refer the reader to the review in \cite{Piller:review}. However, since we use Eq.~(\ref{eq:master2})
 to evaluate nPDFs at some input scale for QCD evolution, in order to provide sensible results
 after the evolution, we should have a reasonable estimate of nPDFs for all $x$. 
We adopted the picture of nuclear modification of PDFs developed in \cite{FS:PRep160,Liuti},
 which suggests that antiquarks in nuclei are not enhanced and the gluons are antishadowed 
and which uses the constraints based on the baryon and energy-momentum
conservation sum rules.
 In our case, like in \cite{FGMS02}, we model the enhancement of the gluon nPDF in the interval
 $0.03 \leq x \leq 0.2$ with a simple function $a(0.2-x)(x-0.03)$ and choose the free coefficient
 $a$ by requiring the conservation of the momentum sum rule for nPDFs. For instance 
for $^{40}$Ca, this requirement gives $a \approx 30$ and about 2-3\% enhancement of the fraction
 of the total momentum of the nucleus carried by the gluons, in accord with the analysis of \cite{FS:PRep160}.

\section{Leading twist nPDFs and structure functions} 
\label{sec:results}

\begin{figure}
\includegraphics[width=13cm,height=13cm]{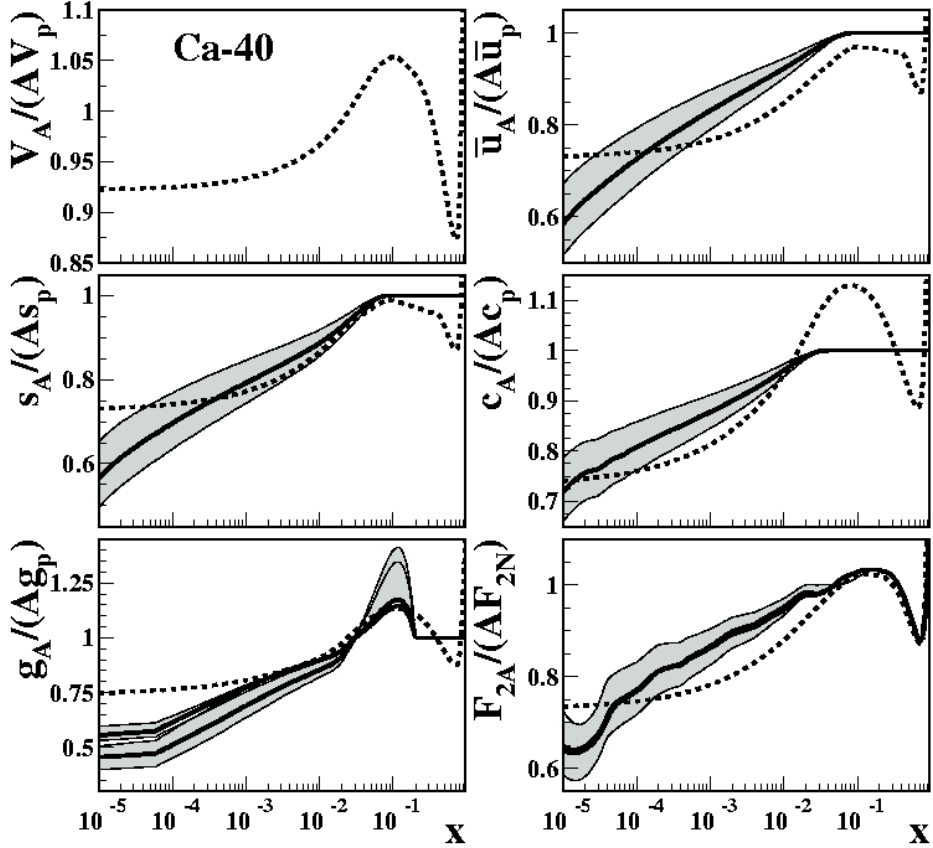}
\caption{The ratio of nuclear to proton NLO parton distributions and the nuclear to  free nucleon
  inclusive structure functions $F_2$ in $^{40}$Ca at $Q=2$ GeV. The leading twist theory results
 (solid curves and the corresponded shaded error bands) are compared to the LO predictions
 by Eskola {\it et al.} \protect\cite{Eskola} (dashed curves).}
\label{fig:input}
\end{figure}

The master equation~(\ref{eq:master2}) allows one to determine NLO  nPDFs at the input scale
$Q_0^2=4$ GeV$^2$.
As an example of such a calculation, we present
ratios of the nuclear (Ca-40) to free proton PDFs and
the ratio of the nuclear to the free nucleon
 structure function, $F_2^N=(F_2^p+F_2^n)/2$,  at $Q^2=4$ GeV$^2$ by
solid curves in Fig.~\ref{fig:input}.
The shaded error bands around the solid curves indicate the uncertainty of the predictions.
For comparison, 
LO predictions 
of Eskola {\it et al.} \cite{Eskola} 
(based on the LO fit to the DIS and Drell-Yan nuclear data) 
for the corresponding ratios are given by the dashed curves.

Note that for the gluon ratio, we give two predictions corresponding to two versions
of the diffractive slope $B_g$ discussed earlier.
Also,
since we do not predict nuclear shadowing for the valence quarks, this information 
should be taken from elsewhere. In our analysis, we use the parameterization by Eskola {\it et al.}
 \cite{Eskola} (see the upper left panel).

 For the parameterization of the proton PDFs, we used the NLO fit CTEQ5M \cite{CTEQ5}.

\begin{figure}
\includegraphics[width=13cm,height=12cm]{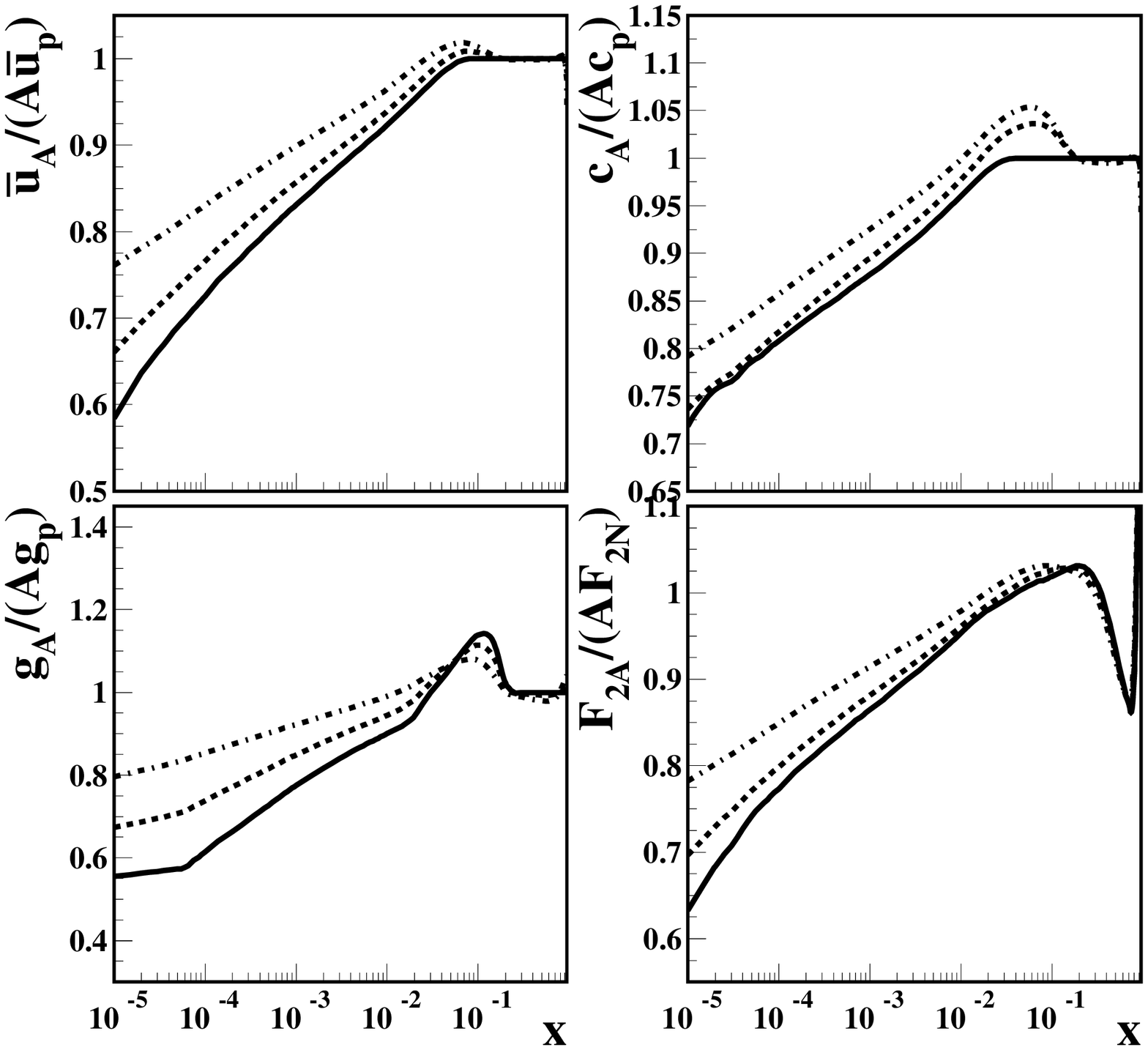}
\caption{Scaling violations for nuclear PDFs in $^{40}$Ca. The solid curves correspond to
$Q^2=4$ GeV$^2$; the dashed curves correspond to $Q^2=10$ GeV$^2$; the dot-dashed curves
correspond to $Q^2=100$ GeV$^2$.}
\label{fig:evo}
\end{figure}

Figure~\ref{fig:evo} presents the $Q^2$-evolution of the ratios in Fig.~\ref{fig:input}. The solid curves 
correspond to $Q^2=4$ GeV$^2$; the dashed curves correspond to $Q^2=10$ GeV$^2$; the dot-dashed curves
correspond to $Q^2=100$ GeV$^2$. The leading twist character of the predicted nuclear shadowing is 
apparent from this figure: The shadowing correction decreases slowly with increasing $Q^2$ and
there is still rather significant nuclear shadowing at  $Q^2=100$ GeV$^2$. 

One can see from Fig.~\ref{fig:input} that our predictions at the
lowest values of Bjorken $x$ significantly differ from those by
Eskola {\it et al.} However, one should  keep in mind that we make
our predictions to the  NLO accuracy, while the fitting to the nuclear 
DIS data in \cite{Eskola} is 
done to the LO accuracy.
One should also note that the parameterization of Eskola
 {\it et al.}~\cite{Eskola} 
assumes that at small $x$, the ratios $F_2^A/(A F_2^N)$ and $g_A/(A g_N)$
 become equal and stay  constant (saturate).
We point out that:
\begin{itemize}
{\item Figure~\ref{fig:input} presents our predictions for the shapes of the nPDFS for $^{40}$Ca,
 which are to be used
 as an input for QCD evolution at the scale $Q_0=2$ GeV. This choice of $Q_0$
is motivated by the fact that
 the 1994 H1 diffractive data has $Q^2 \geq 4.5$ GeV$^2$ and the QCD fit to the data of 
\cite{H1:parametrization} starts at $Q^2=3$ GeV$^2$.
Results of such evolution are presented in Fig.~\ref{fig:evo}.
}
{\item Leading twist theory predicts much more significant nuclear shadowing for quarks and 
gluons than the fits to the fixed-target data of Eskola
 {\it et al.}.~\cite{Eskola}. The latter
{\it assumed} that  shadowing saturates for small $x$, e.g. for 
$x \lesssim 3 \times 10^{-3}$ for  $^{40}$Ca, 
and that higher twist effects are negligible.}
{\item Nuclear shadowing for the gluons is larger than for the quarks 
due to the dominance of gluons in diffractive pdfs (this was further confirmed
 by the recent ZEUS data \cite{ZEUS2004}.
}
{\item Within our model of antishadowing for the gluons by the simple function $a(0.2-x)(x-0.03)$, 
significant variations of the parameter $a$ still lead to the conservation 
(with accuracy better than 1\%) of the parton momentum sum rule.
 Hence, the amount of
 antishadowing for the gluons is not sensitive to the low-$x$ behavior of 
the gluons.} 
{\item Should we compare our predictions to those by Hirai, Kumano
 and Miyama or to the updated fit by Hirai, Kumano and Nagai \cite{Kumano},
 the disagreement in the shadowing predictions, especially for
the gluons, would be much larger. 
For the comparison of the parameterizations of \cite{Eskola} and the first 
of \cite{Kumano}, one can consult \cite{Eskola:last}.
}
{\item Comparing to the parameterization suggested in the work of Li and Wang \cite{HIJING},
 we again find a strong disagreement in the quark channel and, a surprisingly good agreement for the gluons.
 However, the  
parameterization
of \cite{HIJING} are not based on the detailed comparison to all available 
fixed-target data, but rather on the need to fit the RHIC data within the HIJING model.
Note also that this parameterization does not include the enhancement either
 in the quark or gluon channels and, hence, violates the exact QCD momentum
 sum rule for the parton densities.
}
{\item The only NLO QCD fit to nuclear DIS data by de Florian and
  Sassot \cite{deFlorian} produces
 very small nuclear shadowing for nPDFs, which is inconsistent with
 our predictions as well
as with the predictions of \cite{Eskola,Kumano,HIJING}.}
{\item Our predictions for nPDFs and the structure function $F_2^A$ for the nuclei of $^{12}$C, 
$^{40}$Ca, $^{110}$Pd, $^{197}$Au and $^{206}$Pb and for the kinematic range $10^{-5} \leq x \leq 1$ and
$4 \leq Q^2 \leq 10,000$ GeV$^2$ have been tabulated. 
They are available in the form of a simple Fortran
program from V. Guzey upon request, vadim.guzey@tp2.rub.de, 
or from V. Guzey's web page, http://www.tp2.rub.de/$\sim$vadimg/index.html.
}
\end{itemize}

\begin{figure}
\includegraphics[width=12cm,height=12cm]{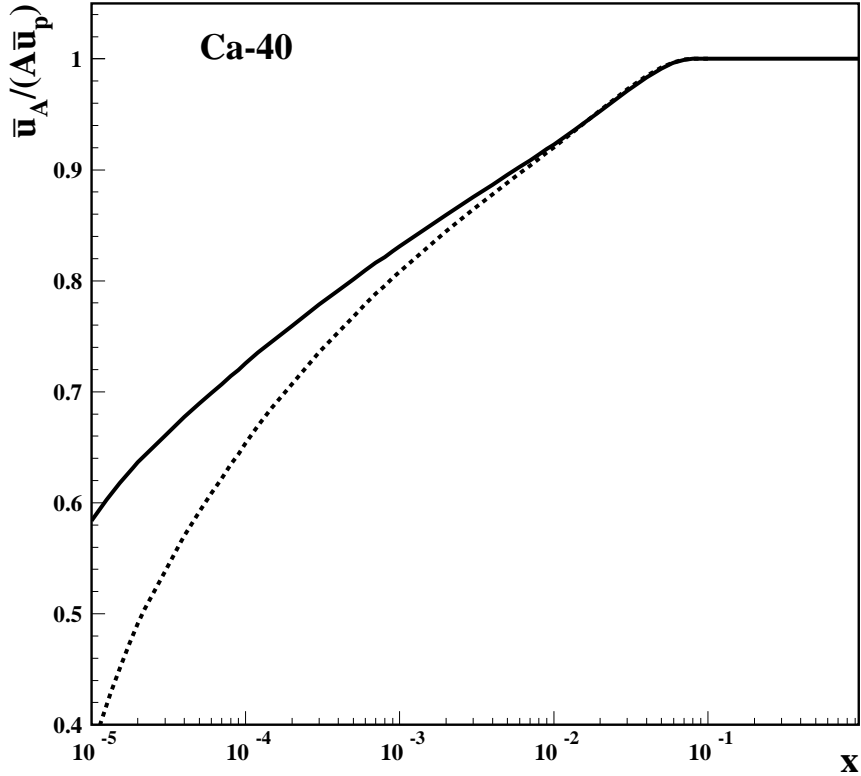}
\caption{The $\bar{u}_A/(A \bar{u}_N)$ ratio for $^{40}$Ca at $Q=2$ GeV.
The solid curve is the result of the full calculation; the dashed curve is 
obtained by neglecting multiple rescatterings.} 
\label{fig:rescattering}
\end{figure}

We also study the importance of the effect of multiple rescatterings, which
is described by the attenuation factor $T(b)$ in  Eq.~(\ref{eq:master2}).
Figure~\ref{fig:rescattering} compares the result of the full calculation 
of the  $\bar{u}_A/(A \bar{u}_N)$ ratio for $^{40}$Ca at $Q=2$ GeV 
(solid curve) with the calculation, when the rescattering effect 
was ignored (dashed curve), i.e. $T(b)$ was set to one
 in  Eq.~(\ref{eq:master2}).
As seen from Fig.~\ref{fig:rescattering}, the rescattering effect becomes
unimportant for $x > 0.005$, i.e. in the kinematics of the fixed-target nuclear
DIS experiments.

Nuclear shadowing corrections to nPDFs become significantly larger,
 when one considers the interactions
 with the target nucleus at small impact parameters. Indeed, since the density of nucleons in
 the center of the nucleus is larger than the average nucleon density, choosing small impact parameters
 corresponds to the increase of the number of scattering centers. 
Introducing the impact parameter dependent nPDFs,
 $f_{j/A}(x,Q^2,b)$, as was done in \cite{FGMS02}
\begin{equation}
\int d^2 b f_{j/A}(x,Q^2,b)=f_{j/A}(x,Q^2) \,,
\label{eq:impact}
\end{equation}
the nuclear shadowing correction to the impact parameter dependent nPDFs
 can be readily 
found from Eq.~(\ref{eq:master2}) by simply removing the integration over 
the impact parameter $b$
\begin{eqnarray}
&&\delta f_{j/A}(x,Q^2,b)=\frac{A(A-1)}{2} 16 B_j \pi {\cal R}e \Bigg[\frac{(1-i\eta)^2}{1+\eta^2} 
 \int^{\infty}_{-\infty} dz_1 \int^{\infty}_{z_1} dz_2 \int^{x_{\Pomeron,0}}_{x} d x_{\Pomeron}  
\nonumber\\
&&\times f_{j/N}^{D(3)}(\beta,Q^2,x_{\Pomeron}) \rho_A(b,z_1) \rho_A(b,z_2) e^{i x_{\Pomeron} m_N (z_1-z_2)} e^{-(A/2)(1-i\eta)\sigma_{{\rm eff}}^j \int_{z_1}^{z_2} dz \rho_A(b,z)} \Bigg] \,.
\label{eq:master3} 
\end{eqnarray}
  
\begin{figure}
\includegraphics[width=13cm,height=13cm]{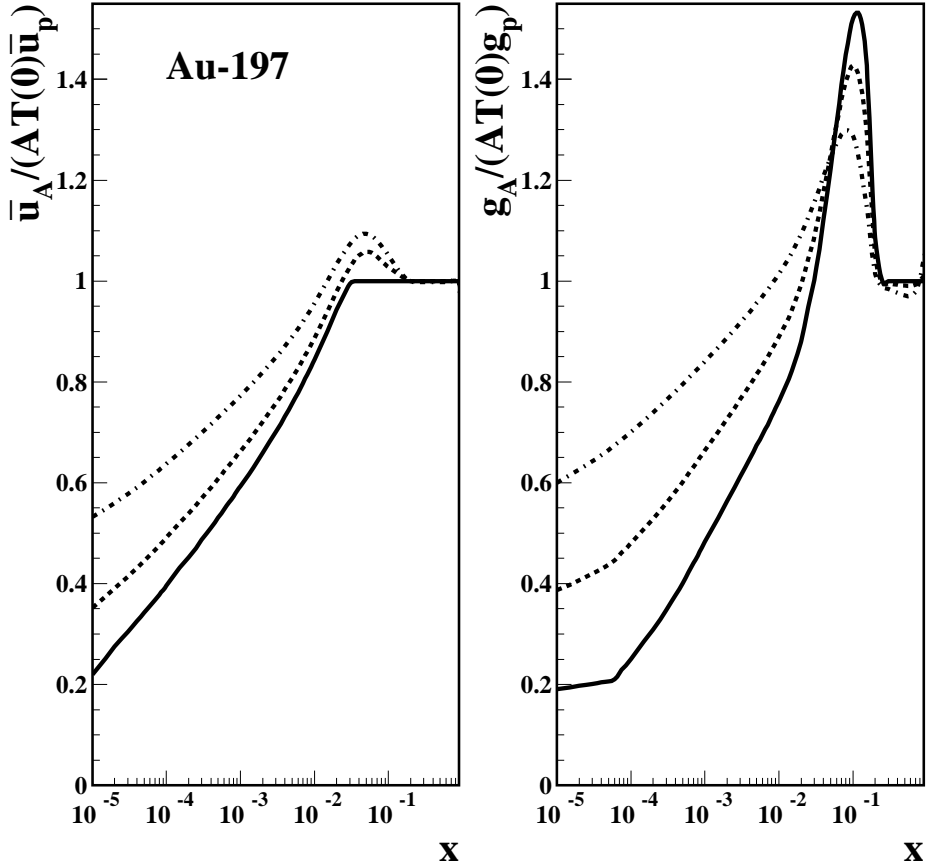}
\caption{Nuclear shadowing at zero impact parameter: The ratios 
${\bar u}_A(x,Q^2,0)/(A T(0){\bar u}_N(x,Q^2,0))$ and  
$g_A(x,Q^2,0)/(A T(0) g_N(x,Q^2,0))$ for $^{197}$Au at $Q^2=4$ GeV$^2$ (solid), $Q^2=10$ GeV$^2$ (dashed) and
$Q^2=100$ GeV$^2$ (dot-dashed).}
\label{fig:central}
\end{figure}

The results of the evaluation of the nuclear shadowing correction using Eq.~(\ref{eq:master3}) at
 the zero impact parameter for anti $u$-quarks and gluons in $^{197}$Au
 are presented in Fig.~\ref{fig:central} in terms of the ratios ${\bar u}_A(x,Q^2,0)/(A T(0){\bar u}_N(x,Q^2,0))$ and  
$g_A(x,Q^2,0)/(A T(0) g_N(x,Q^2,0))$.
The solid curves correspond to
$Q^2=4$ GeV$^2$; the dashed curves correspond to $Q^2=10$ GeV$^2$; the dot-dashed curves
correspond to $Q^2=100$ GeV$^2$.
Note that the factor $T(0)=\int dz \rho_A(b=0,z)$ provides the correct normalization of the impulse 
approximation term, see \cite{FGMS02} for details. 
The impact parameter-dependent nPDFs have been tabulated and are available upon request from V. Guzey or from the following web cite, http://www.tp2.rub.de/$\sim$vadimg/index.html.
They were already used by R.~Vogt in the analysis of the $J/\psi$
 production at RHIC and
were found to be in a reasonable agreement with the data \cite{Vogt}.

\section{Comparison to the data and evidence for higher twist effects}
\label{sec:ht}

\begin{figure}
\includegraphics[width=13cm,height=13cm]{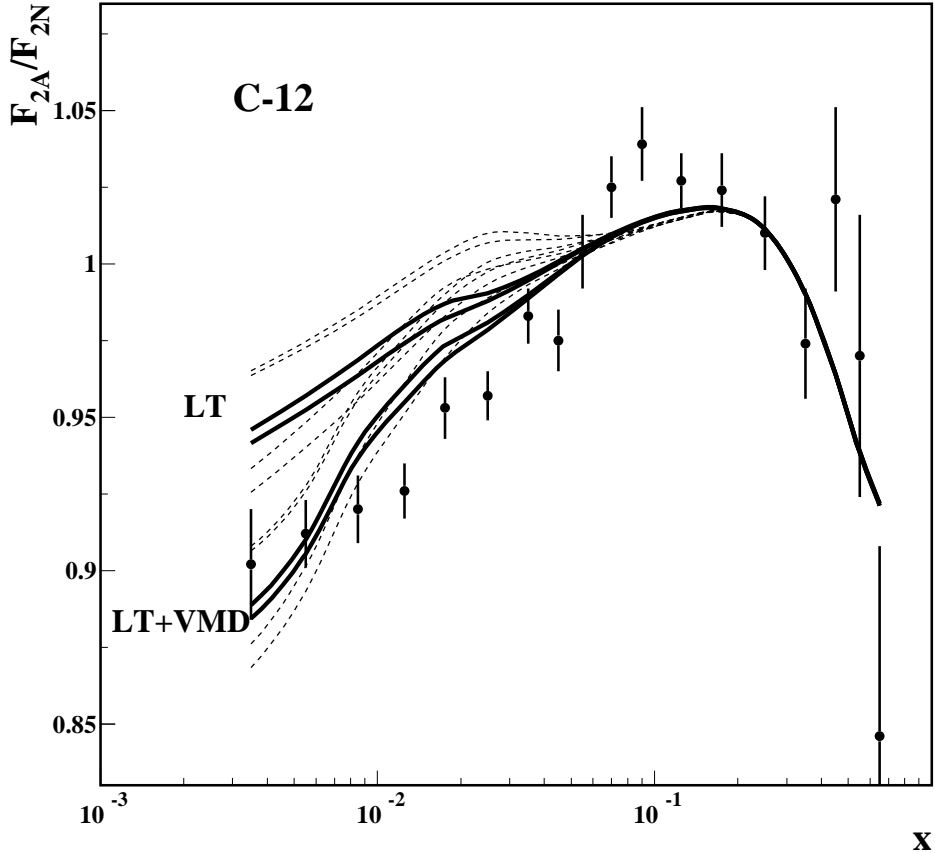}
\caption{Comparison of the leading twist theory results 
(upper set of 
solid curves and associated dashed error bands)
to the NMC data on $F_2^{Ca}/F_2^N$  \protect\cite{NMC1}. 
The lower set of the solid curves is obtained by adding the VMD contribution
using Eq.~(\ref{eq:vmd}).
}
\label{fig:nmcc12}
\end{figure}

\begin{figure}
\includegraphics[width=13cm,height=13cm]{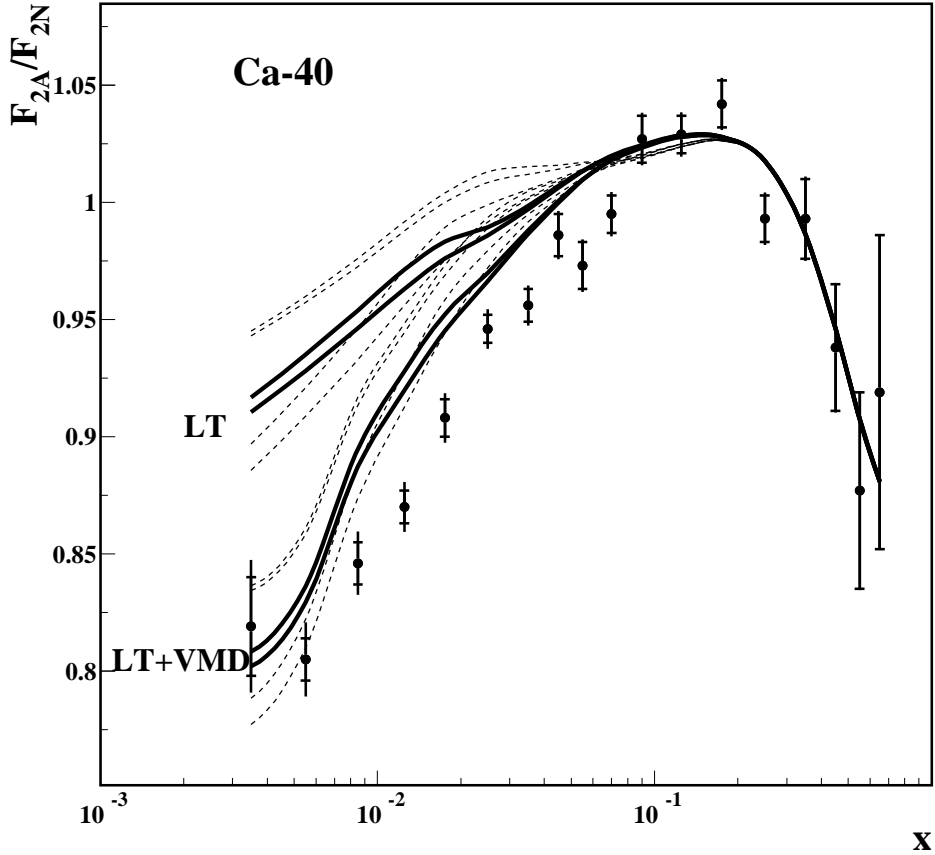}
\caption{Comparison of the leading twist theory results (upper set of 
solid curves and associated dashed error bands)
to the NMC data on $F_2^{Ca}/F_2^N$ \protect\cite{NMC1}.
The lower set of the solid curves is obtained by adding the VMD contribution
using Eq.~(\ref{eq:vmd}).
}
\label{fig:nmcca40}
\end{figure}

\begin{figure}
\includegraphics[width=13cm,height=13cm]{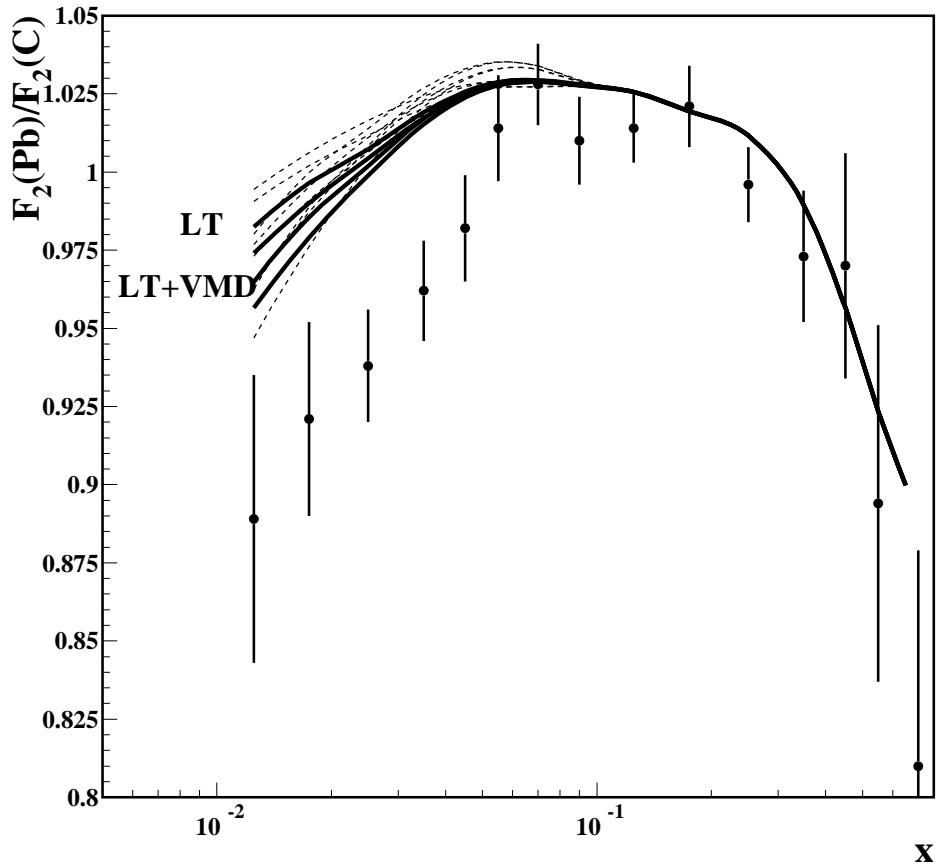}
\caption{Comparison of the leading twist theory results 
(upper set of 
solid curves and associated dashed error bands)
to the NMC data on  $F_2^{Pb}/F_2^{C}$ \protect\cite{NMC2}.
The lower set of the solid curves is obtained by adding the VMD contribution
using Eq.~(\ref{eq:vmd}).
}
\label{fig:nmcpb}
\end{figure}

Our predictions for the $F_2^A/(AF_2^N)$ ratio, where $F_2^N=(F_2^p+F_2^n)/2$, can  be compared to the
NMC data \cite{NMC1,NMC2}. However, since the low-$x$ data points correspond
 to low $Q^2$, we cannot make a direct
 comparison with those points. Therefore, we simply evaluate
 $F_2^A/(AF_2^N)$ at
 $Q^2=Q_0^2 = 4$ GeV$^2$ for the data points with $Q^2 < 4$ GeV$^2$.
Figures \ref{fig:nmcc12} and \ref{fig:nmcca40} compare predictions
of our leading twist model (upper set of solid curves with the associated error
bands denoted by dashed curves)  
to the NMC data on $^{12}$C and $^{40}$Ca \cite{NMC1}; 
Fig.~\ref{fig:nmcpb} makes a comparison to the NMC $F_2^{Pb}/(F_2^C)$
 ratio \cite{NMC2}.

One can see from Figs.~\ref{fig:nmcc12}, \ref{fig:nmcca40} and  \ref{fig:nmcpb} that the agreement between the data points and our calculations
 at low $x$ is poor.
Regardless the fact that our model predicts a significant nuclear shadowing
effect for low-$x$, $x < 10^{-3}$, see Fig.~\ref{fig:input},
 nuclear shadowing rather rapidly decreases
when $x$ approaches the values of $x$ probed in fixed-target nuclear DIS 
experiments.
Of course, one might argue that we are comparing our predictions at $Q^2=4$ GeV$^2$ 
 to the data with much lower  $Q^2$ values. 
For instance,  in Fig.~\ref{fig:nmcca40} for the first five data points, 
the average values of $Q^2$ are $\langle Q^2 \rangle=(0.60, 0.94, 1.4, 1.9, 2.5)$ GeV$^2$.
We have explicitly checked that the backward QCD evolution of our predictions
 down to $Q^2=2$ GeV$^2$
changes the predictions only a little.
Therefore, since our approach to nuclear 
shadowing includes the entire leading twist contribution to the nuclear shadowing correction 
(one should keep in mind a significant uncertainty due to the 
unaccounted Reggeon contribution, see Appendix~\ref{sec:reggeon}),
 the disagreement with the NMC low-$x$ data compels us
 to conclude that {\it the low-$x$ NMC data \cite{NMC1,NMC2} 
could
contain significant higher twist effects,
 which contribute approximately 50\% to  the nuclear shadowing correction  to  $F_2^A$.}

At the same time, the $A$-dependence is reproduced reasonably well (see below)
indicating that the inadequate modeling of the diffraction at low $Q^2 $ and 
$x$ is to blame.
Indeed,
 it is very natural to have rather significant higher twist
effects at small $Q^2$ since, for this kinematics, the contribution of 
small diffractive masses $M_X$ becomes important. Production of small 
diffractive masses $M_X$ is dominated by the production of vector mesons,
 which is definitely a higher twist phenomenon. In this kinematics, 
the leading twist H1 parameterization of diffraction \cite{H1:parametrization}
 underestimates by approximately factor two the diffractive cross section
 as illustrated by the following estimate.

Using the definition of the diffractive differential cross section in
 terms of the diffractive structure function $F_2^{D(3)}$ \cite{H1:1994} and
 the relation between the differential cross section on the lepton level
 to the total cross section on the virtual photon level
 (the Hund convention for the virtual photon flux), 
$\sigma(\gamma^{\ast}p \to X p)$, the latter can be written as
\begin{equation}
\sigma(\gamma^{\ast}p \to X p)=\frac{4 \pi^2 \alpha_{e.m.}}{Q^2} \int_x^{x_{\Pomeron,0}} dx_{\Pomeron} F_2^{D(3)}(\beta,Q^2,x_{\Pomeron}) \,.
\label{eq:rho1}
\end{equation}
Then, if we restrict the integration in Eq.~(\ref{eq:rho1}) by low
 diffractive masses $M_X$, $M_X \leq 1$ GeV,
 the resulting $\sigma(\gamma^{\ast}p \to X p)$ can be compared to the
 cross section of electroproduction of vector mesons (dominated by the $\rho$ meson).
 For instance,
a comparison to the HERMES data on exclusive leptoproduction of $\rho^{0}$
 mesons from hydrogen \cite{HERMES:rho} at low $Q^2$ and $W$,
 $\langle Q^2 \rangle=0.83$ GeV$^2$ and $\langle W \rangle=5.4$ GeV,
 demonstrates that the calculation using Eq.~(\ref{eq:rho1}) 
(with the restriction $M_X \leq 1$ GeV)
 gives only 40\% of the experimental 
value $\sigma(\gamma^{\ast}p \to \rho^{0} p)=2.04 \pm 0.10 \pm 0.43$ $\mu$b.
This observation means that, in the considered case, there is no duality
between the continuum and resonance contributions to low-mass inclusive
diffraction.

In order to quantitatively study our conclusion about the significant role
 of the higher twist contribution to nuclear shadowing, we explicitly add the
contribution of $\rho$, $\phi$ and $\omega$ vector mesons to our leading
twist predictions in the spirit of the vector meson dominance (VMD) model.
 However, since about 50\% of the vector meson contribution
is already contained in the parameterization of inclusive diffraction,
we weigh the contribution of the vector mesons to nuclear shadowing by the 
factor $1/2$ (this enables us not to double-count the vector meson
contribution). Therefore, the VMD contribution to the shadowing correction 
to the structure function $F_2^A$ reads \cite{BPY,Ratzka}
\begin{eqnarray}
&&\frac{\delta F_{2\,A}^{VMD}}{A F_2^N}=-\frac{1}{2} \frac{A-1}{2}\frac{Q^2 (1-x)}{\pi F_2^N} \sum_{V=\rho,\phi,\omega} \frac{\sigma_V^2(Q^2)}{f_V^2} \left(\frac{m_V^2}{Q^2+m_V^2}\right)^2 \nonumber\\
&& \times \int d^2 b \int^{\infty}_{-\infty} dz_1 \int^{\infty}_{z_1} dz_2  
 \rho_A(b,z_1) \rho_A(b,z_2) \cos\left(\Delta_V(z_2-z_1)\right)e^{-(A/2)\sigma_{V} \int_{z_1}^{z_2} dz \rho_A(b,z)}  \,,
\label{eq:vmd}
\end {eqnarray}
where $\Delta_V=x \, m_V (1+m_V^2/Q^2)$. The VMD parameters $\sigma_V$
and $g_V$ assume their usual values \cite{Ratzka}. For the nucleon structure 
function $F_2^N$ for low $Q^2$ and $x$, we used the NMC parameterization 
\cite{NMC:F2N} for $x > 0.006$ and the ALLM fit \cite{ALLM} for $x < 0.006$.
In addition, since the values of $Q^2$, where expression~(\ref{eq:vmd}) is
used can be as large as 10 GeV$^2$, we take into account the effect of the
vector meson size decrease by introducing the explicit $Q^2$ dependence of
$\sigma_V$: $\sigma_V$ decreases by the factor of 4 when $Q^2$ increases from
$0.6-0.7$ GeV$^2$ to 10 GeV$^2$ \cite{Koepf}.
 
The inclusion of the VMD contribution to the shadowing correction dramatically
improves the agreement between our calculations and the NMC data for
the considered cases of $^{12}$C and $^{40}$Ca, see the lower set of the
solid curves in Figs.~\ref{fig:nmcc12} and \ref{fig:nmcca40}.
This  supports our conclusion about the 50\% contribution
of higher twist effects to nuclear shadowing in the fixed-target
nuclear DIS kinematics.

As seen from Fig.~\ref{fig:nmcpb}, the inclusion of the VMD contribution
 does not significantly improve the agreement between our calculations and 
the data. Since the data points have rather large $Q^2$, $Q^2 \geq 3.4$ 
GeV$^2$, it is natural that the contribution of the $\rho$, $\omega$ and
 $\phi$ mesons is rather small. 
 
 At the same time, the amount  of shadowing is sensitive to the 
diffractive masses
up to $Q \sim 1.7$ GeV,  where an enhancement is also possible as compared
 to the leading twist fit that we employ. One has to emphasize that one is
 dealing here with $x_{\Pomeron} > 0.01$, where the Pomeron component of the
 diffractive PDFs
is known to underestimate the data by a large factor, for the recent discussion 
see \cite{ZEUS2004}.  
 This contribution is usually referred to as the Reggeon contribution to 
 diffractive PDFs. However, 
 uncertainties in the theoretical treatment of this contribution are very large due to 
possible 
interference between the Pomeron
and Reggeon contribution, 
unreliable parameter $\eta$ for the
Reggeon contribution, difficulty to reliably extract the Reggeon contribution
from the diffractive data.
Hence we were forced to neglect this contribution in our numerical analysis.
Therefore, the theoretical uncertainty due to the Reggeon contribution for
$x > 0.01$ is almost out of control and, hence, our
calculation  for $x > 0.01$ become much less reliable. 
We refer the reader to Appendix~\ref{sec:reggeon} where we examine
the Reggeon contribution and its influence of our predictions of nuclear
shadowing and on the comparison of our results to the NMC nuclear DIS data. 
The domain, where the presented leading twist model is best justified, is 
$x < 5 \times 10^{-3}$ and $Q^2 \geq 4$ GeV$^2$.

It is worth emphasizing here  that our conclusion about the importance of 
the higher twist (HT) effects relies on the existing 
parameterizations  of the leading twist (LT) diffractive PDFs which exclude from the
 fits the region of small diffractive masses. Hence the large $\beta $ 
region is, to the large extent, an extrapolation of lower $\beta$ data or
 effectively due to backward evolution of large $\beta$ and high $Q^2$ data,
 which is  rather not stable. So one cannot 
a priory exclude that there exists a LT parameterization which satisfies 
a local duality requirement  in the large $\beta$, low $Q^2$ region. An investigation
 of such a possibility would require studying diffraction at lower energies
 than presently accessible at HERA.

\begin{figure}
\includegraphics[width=13cm,height=13cm]{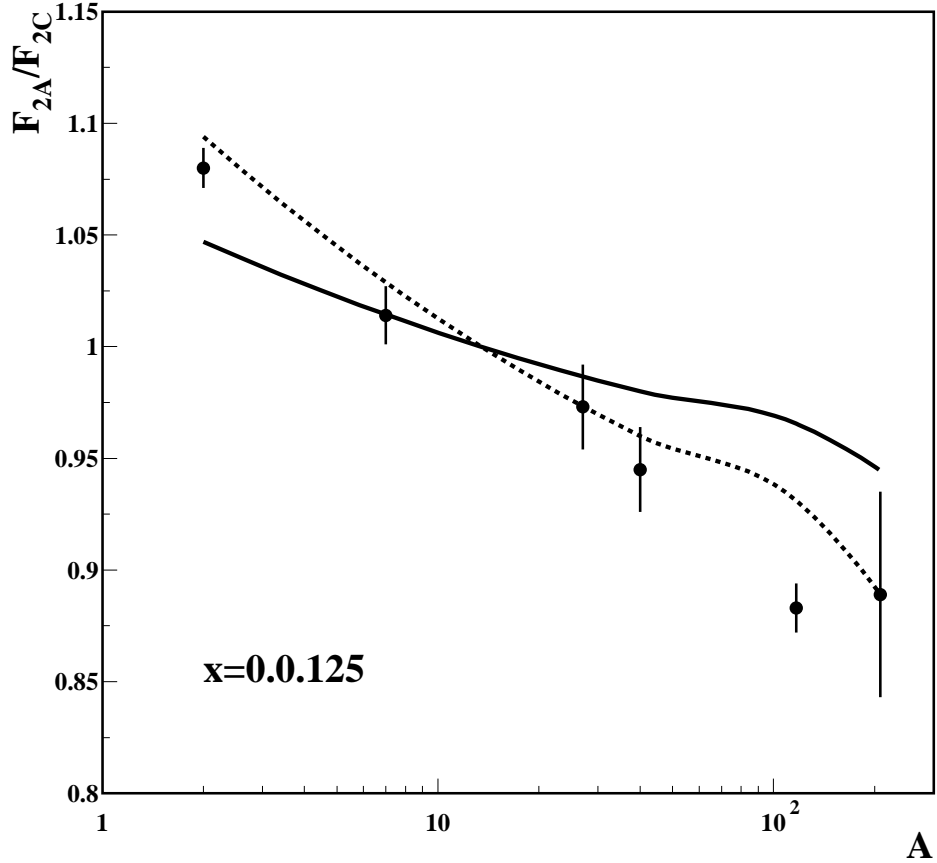}
\caption{$A$-dependence of nuclear shadowing.
The NMC data on $F_2^A/F_2^C$ \protect\cite{NMC1,NMC2} are compared to 
our
LT+VDM
 predictions (solid curve). The dashed curve is obtained by scaling up
the shadowing correction of the solid curve by the factor two.
}
\label{fig:adep}
\end{figure}

One
 can also study the $A$-dependence of nuclear shadowing.
Figure~\ref{fig:adep} presents a comparison of the NMC data on $F_2^A/F_2^C$
\cite{NMC1,NMC2} to our model calculations. All data points correspond to
$x=0.0125$; the $F_2^D/F_2^C$ data point has $Q^2=2.3$ GeV$^2$ and all other
data points have $Q^2=3.4$ GeV$^2$.
The solid curve is our main prediction, which is obtained as a sum of the
leading twist shadowing and the VMD contribution evaluated using
Eq.~(\ref{eq:vmd}). The dashed curve is obtained by increasing the nuclear
shadowing correction of the solid curve by the factor two.

As seen from Fig.~\ref{fig:adep}, our calculation reproduces fairly well the $A$-dependence of nuclear shadowing. However, as was discussed previously in connection with Fig.~\ref{fig:nmcpb}, even with the higher twist 
VMD contribution included, our calculation
 systematically underestimates the absolute value of the shadowing effect.
For the heavier nuclei,
we should have had
a 2-3 times larger nuclear shadowing, as indicated by the dashed curve.
The reason for the discrepancy between our calculations and the data is a
significant Reggeon contribution to hard diffraction at $x=0.0125$, which 
we neglect in our 
calculation.
 Because of this, the theoretical uncertainty of our
results becomes very large for $x > 0.01$.

\begin{figure}
\includegraphics[width=13cm,height=13cm]{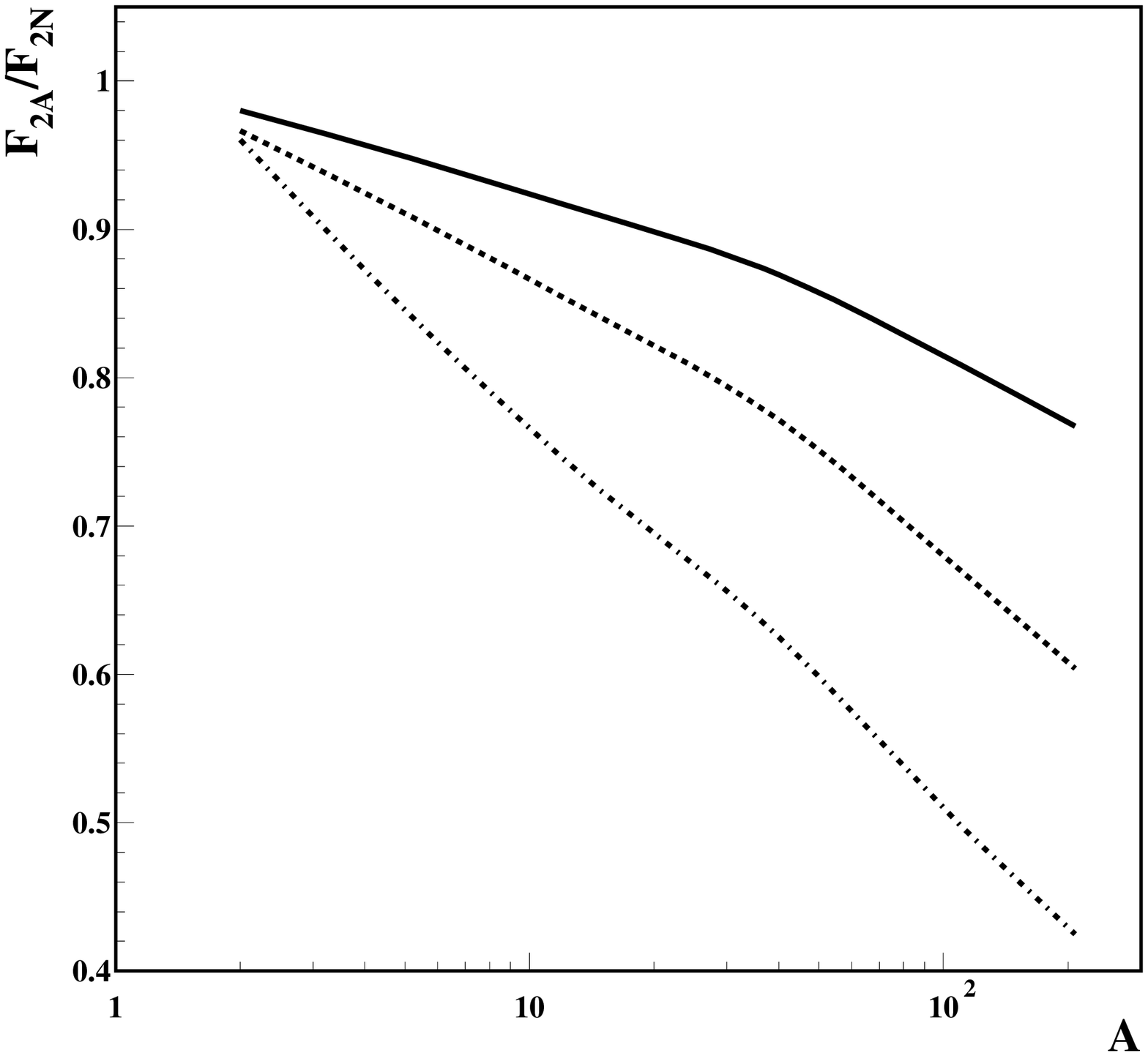}
\caption{The $A$-dependence of nuclear shadowing at $Q^2 = 4$ GeV$^2$.
The solid curve corresponds to $x=10^{-3}$;
the dashed curve corresponds to $x=10^{-4}$;
the dot-dashed curve corresponds to $x=10^{-5}$.
}
\label{fig:adep_lowx}
\end{figure}

The leading twist model of nuclear shadowing works best in the kinematics
where the VMD and Reggeon contributions
to hard diffraction
 are small corrections and where
the data on hard diffraction at HERA were taken, i.e. for $Q^2 \geq 4.5$ 
GeV$^2$ and $x \leq 5 \times 10^{-3}$. An example of the proper
application of the leading twist model of nuclear shadowing is
presented in Fig.~\ref{fig:adep_lowx}, which depicts the $A$-dependence of
$F_2^A/(A F_2^N)$ at $Q^2 = 4$ GeV$^2$ for $x=10^{-3}$ (solid curve),
$x=10^{-4}$ (dashed curve) and $x=10^{-5}$ (dot-dashed curve).

Our conclusion about the importance of the higher twist effects at
 small $x$ and small $Q^2$ in
 the fixed-target data
 is in a broad  agreement with phenomenological approaches to 
nuclear shadowing, which include
 both the scaling (leading twist) and lowest mass ($\rho$, $\omega$ and $\phi$) vector meson
 (higher twist) contributions.  
(The only possible way to avoid such a conclusion would be the local
 duality scenario which we discussed 
earlier in this section.)
 In \cite{Thomas}, the scaling contribution arises as the effect of 
the diffractive scattering, 
quite similarly in the spirit to the present work.
However, it is difficult to assess the comparability of the pre-HERA
 parameterization of diffraction
 used in \cite{Thomas} with the modern HERA data on hard diffraction. 
More importantly, the effect
 of nuclear shadowing was discussed for the nuclear structure function $F_2$ and not for nPDFs
 (this comment also applies to all other work mentioned below).
In other approaches, the scaling contribution results from  the $q
 \bar{q}$ continuum of the virtual photon wave function \cite{Ratzka},
 or from the contributions of higher mass vector mesons
 \cite{Badelek,Shaw}, or from the
aligned  $q \bar q $ jets \cite{FS:PRep160,FS89},
or from the asymmetric $q \bar{q}$ fluctuations of the 
virtual photon \cite{Nikolaev}.
Note also that in the case of the real photon interaction with 
nuclei (the accurate data  on the real 
photon diffraction for
 the relevant energies is available, see \cite{Chapin}), 
 the shadowing data agree well with the Gribov's theory, see the discussion in
Ref.~\cite{Piller:review}.

We would like to point out that a fairly good description of  the NMC 
data was achieved in \cite{Kaidalov}, 
which uses the approach to nuclear shadowing 
based on its relation to diffraction on the nucleon.
 As an input for the their calculation,
 the authors used the phenomenological parameterizations of the inclusive and diffractive structure 
functions of the nucleon, which fit well the inclusive and diffractive data.
 However, in contrast to our
 strictly leading twist analysis, the phenomenological parameterizations of 
 \cite{Kaidalov} 
have the $Q^2$-dependence of the form $(Q^2/(Q^2+a))^b$, where $a$ and $b$ are numerical parameters.
 Hence, the analysis of \cite{Kaidalov} effectively includes higher twist contributions, 
which indirectly confirms our conclusion that a good description of the NMC data~\cite{NMC1,NMC2} 
is impossible to achieve without the inclusion of the higher twist effects (contribution of vector mesons).

In addition, we would like to point to other important difference between 
the present analysis and the
 analysis of  \cite{Kaidalov}.  In order to evaluate nuclear shadowing as
 a function of $Q^2$,
 the authors of  \cite{Kaidalov} apply an equation similar in the spirit to our Eq.~(\ref{eq:master2}) at all $Q^2$.
 As we explain in the end of Sect.~\ref{sec:theory}, the application 
of Eq.~(\ref{eq:master2})
 at large $Q^2$ violates QCD evolution because one then ignores
 the proper increase  of the fluctuations of 
$\sigma_{{\rm eff}}^j$ as a result of the QCD evolution.
Also, neglecting proper QCD evolution, one neglects the contribution
 of larger $x$ effects 
-- antishadowing and EMC effects -- to the small-$x$ region.
The second major difference is that the use of the QCD factorization
 theorem for hard diffraction allowed us
to make predictions for nPDFs. Since this factorization theorem is not used in \cite{Kaidalov}, only
nuclear structure function $F_2^A$ is considered.
Finally, the well-understood and important 
effect of the decrease of the coherence length
with the increase of $x$, i.e. the factor $e^{i x_{\Pomeron} m_N (z_1-z_2)}$, 
was ignored in \cite{Kaidalov}.

\begin{figure}
\includegraphics[width=12cm,height=12cm]{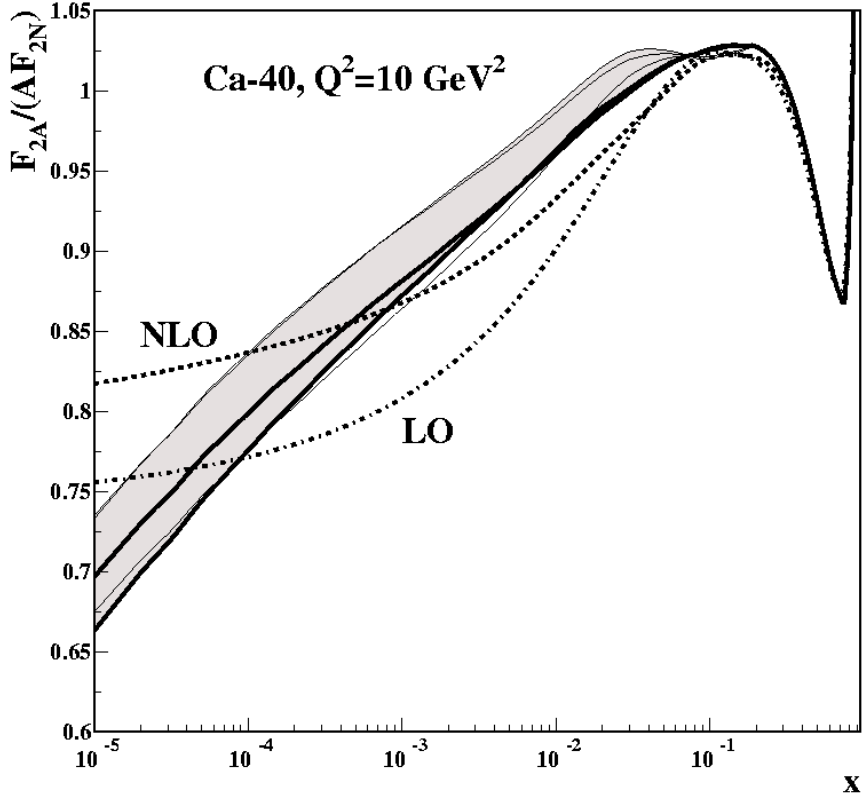}
\caption{LO vs. NLO evolution. The LO fits for nuclear PDFs of Eskola {\it et al.} \protect\cite{Eskola}
are evolved to $Q^2=10$ GeV$^2$ to LO (dot-dashed curve) and to NLO (dashed curve) accuracy.
Our leading twist nuclear shadowing predictions are given by the solid curves and the 
associated error bands.}
\label{fig:f2eskola}
\end{figure}

Next we 
discuss the importance of the 
next-to-leading order (NLO) effects in the nuclear structure function $F_2$.
Using the LO parameterization for the ratios of the nuclear to proton PDFs 
of Eskola {\it et al.} \cite{Eskola} 
 at the initial scale $Q_0=1.5$ GeV, we perform  QCD evolution to $Q^2=10$ GeV$^2$ both 
with NLO and LO accuracy. 
The resulting $F_2^{Ca}/(A F_2^N)$  ratios after the NLO evolution (dashed curve) and LO evolution
 (dot-dashed curve) are presented in Fig.~\ref{fig:f2eskola}. 
For the proton PDFs, we use CTEQ5 parameterizations~\cite{CTEQ5}:  CTEQ5M
for the NLO calculations and CTEQ5L for the LO calculations.
For comparison, we also present $F_2^{Ca}/(A F_2^N)$ (solid curves and the associated error bands)
calculated using our leading twist model. The two solid curves correspond to the two scenarios of
nuclear shadowing for gluons (two models for $B_g$).
A significant difference between the solid and dashed curves in Fig.~\ref{fig:f2eskola}
 demonstrates that the effects associated with the NLO QCD evolution and NLO expression for the structure
 function $F_2$ are important both in the very low-$x$ region and in the $x$-region of the fixed-target data,
 $x > 0.003$. This gives us another indication that the LO fits to the fixed-target data of 
\cite{Eskola,Kumano,HIJING} must have significant
 intrinsic uncertainties, especially at low-$x$, 
where there is no data and the fits are extrapolations.
In addition, Fig.~\ref{fig:f2eskola} demonstrates that it is not self-consistent to use the LO fits for 
nPDFs in the NLO calculations of various hard processes with nuclei, which require NLO nPDFs as an input.

\section{Extraction of the neutron $F_{2n}$}
\label{sec:f2n}

The determination of the small-$x$ behavior of the valence quark distributions
 relies heavily on the use of the $F_2^D$ and $F_{2p}$ data for the 
determination of  the $F_{2p}-F_{2n}$ difference.
The main problem is that at $x \le 0.03$, the difference is comparable or
even smaller than the nuclear shadowing correction for the deuteron, which we
denote as $\delta(x,Q^2)$,
 $\delta(x,Q^2) \equiv 1-
F_2^D(x, Q^2)/(F_{2p}(x, Q^2)+ F_{2n}(x, Q^2))$.
Then the $F_{2p}-F_{2n}$ difference reads
\begin{equation}
F_{2p}(x, Q^2) -  F_{2n}(x, Q^2)= 2 F_{2p}(x,Q^2)-F_2^D(x, Q^2)\left(1+\delta(x,Q^2)\right)\,.
\end{equation}

In view of the discussed uncertainties in the model of diffraction (higher twist effects, subleading contributions) in the
 kinematics, where the NMC took their most 
accurate data for $\mu - D$ and $\mu-p$ scattering,
one can hardly use 
the Gribov theory to calculate reliably $\delta(x,Q^2)$.

However,
 the same 
mechanism works both for the deuteron and for heavier nuclei as strongly suggested by Fig.~\ref{fig:adep}. Hence we can employ the information on the 
$F_{2}^A/F_2^D$ ratios
in order to determine the value of $\delta(x,Q^2)$.
Combining the results of the present analysis with the results of 
our analysis of nuclear
shadowing corrections to the deuteron $F_2^D$ \cite{FGS:deuteron},
we find that
\begin{equation}
\frac{1}{7} \left(1- \frac{F_2^{C}(x, Q^2)}{F_2^D(x, Q^2)} \right) \approx  1 - \frac{F_2^D(x, Q^2)}{F_{2p}(x, Q^2)+ F_{2n}(x, Q^2)} \,.
\end{equation}
Using this equation and the NMC nuclear data, we find 
that $\delta(0.0125, 2.3 \, {\rm GeV}^2) \approx 0.011 \pm 0.001$.
 Using the CTEQ5L fit to the nucleon PDFs \cite{CTEQ5}, we observe that for this kinematics nuclear shadowing significantly
changes the $F_{2p}(x, Q^2) -  F_{2n}(x, Q^2)$  difference
\begin{equation}
\frac{F_{2p}(x, Q^2) -  F_{2n}(x, Q^2)}{F_{2p}(x, Q^2) -  \tilde{F}_{2n}(x, Q^2)}=-\delta \frac{1+R}{1-R}=-0.63 \pm 0.06\,,
\end{equation}
where $R=\tilde{F}_{2n}/F_{2p}$; $\tilde{F}_{2n}$ is
the neutron structure function extracted from the deuteron $F_2^D$ ignoring the
shadowing correction.
 In this estimate, 
 we included only errors due to the experimental uncertainty in the 
 $F_2^{C}/F_2^D$ ratio.

\section{Conclusion and Discussions}

The main results of this paper can be summarized as follows:
\begin{itemize}

{\item We explain the derivation of the leading twist theory of nuclear shadowing in
 DIS on nuclei that relates nuclear shadowing in DIS on nuclei to DIS diffraction on the proton.
 The theory  enables us to predict nuclear shadowing for individual nuclear PDFs in a model-independent
 way at small $x$, $10^{-5} \lesssim x \lesssim 10^{-2}$. At larger $x$, 
other nuclear effects (antishadowing,
 EMC effect) and details of the mechanism of diffraction at
 high $x_{\Pomeron}$ introduce a 
large model dependence and uncertainty.}

{\item Nuclear shadowing corrections to nPDFs are found to be large.
 In particular, we predict
 larger shadowing than given by the fits by Eskola {\it et al.} \cite{Eskola} 
for gluons for all $x$ 
and  for quarks for $x < 5 \times 10^{-4}$.
In a stark disagreement with all other approaches, we predict larger 
nuclear shadowing for gluons
than for quarks.
}

{\item The presented formalism is applied to evaluate nuclear shadowing for 
nPDFs at all 
impact parameters. As one decreases the impact parameter, the effect of nuclear
 shadowing increases.}

{\item 
The results of our purely leading twist calculations
 for the $F_2^{C}/F_2^N$, $F_2^{Ca}/F_2^N$ and $F_2^{Pb}/F_2^C$ 
 ratios disagree with the corresponding fixed-target NMC data \cite{NMC1,NMC2} at low $x$ and low
 $Q^2$. While we cannot compare our prediction directly to the data at the
 $Q^2$ values of the first five data points, we notice that the
 backwards QCD evolution is small and it does not seem to increase
 nuclear shadowing. Hence, we conclude that the NMC data with $Q^2 <
 4$ GeV$^2$  
 are likely to contain
  a significant amount (about 50\%)
 of higher twist effects. 
This is supported by the explicit inclusion of the $\rho$, $\omega$ and
$\phi$ meson contributions to the shadowing correction in the spirit
of the vector meson dominance model.  
An alternative scenario would be the existence of  a local duality pattern for
 diffraction at 
$x_{\Pomeron} > 0.01$ where so far no data were taken, see 
Appendix~\ref{sec:reggeon}. 
This
 implies that a
leading twist
 QCD analysis of the low-$x$ and low-$Q^2$ fixed-target data will not
 produce reliable results for the low-$x$
nuclear PDFs.}

{\item  Using general features of the Gribov theory and the data on $A> 2$  
nuclei, it is possible to develop a reliable procedure for the extraction
 of the neutron $F_{2n}$ in the NMC small-$x$ kinematics.}

{\item Our predictions for nPDFs, impact parameter-dependent nPDFs and the
 structure function $F_2^A$
 for the nuclei of $^{12}$C, 
$^{40}$Ca, $^{110}$Pd, $^{197}$Au and $^{206}$Pb and for the kinematic range $10^{-5} \leq x \leq 1$ and
$4 \leq Q^2 \leq 10,000$ GeV$^2$ have been tabulated. They are available in the form of a simple Fortran
program from V. Guzey upon request, vadim.guzey@tp2.rub.de.
The QCD evolution was carried out using the QCDNUM evolution package \cite{QCDNUM}.
}
\end{itemize}

We must mention that there is a renewed interest in nuclear shadowing because
 of the recent
surprising measurements of the suppression of production of hadrons with high
 $p_t$ in deuteron-gold collisions at RHIC. 
It was claimed that the observed suppression is a spectacular confirmation 
of the Colored
Glass Condensate model. However, in the RHIC kinematics nPDFs are probed at
 relatively large
values of Bjorken $x$, on average $x > 0.01$, which is beyond the domain of
 the 
Colored Glass Condensate model. Since leading twist nuclear shadowing is rather weak for $x > 0.01$,
 nuclear shadowing cannot be 
responsible for the dramatic effect of the suppression of the hadron spectra at forward
 rapidities at 
RHIC, see \cite{Vogelsand}.  
However, shadowing in the forward RHIC
kinematics can be observed by 
selecting appropriate two jet production kinematics.

After the first version of this paper was released, there appeared an analysis 
 \cite{Qiuvitev} which calculates higher twist effects 
in shadowing. Similarly to us, the authors come to the conclusion that 
the higher twist effects in the fixed-target kinematics are large. Within
uncertainties
of their analysis, the higher twist effects could be even
responsible for all shadowing observed at fixed target energies.  
So far the connection of the approach of \cite{Qiuvitev} to the Gribov
theory is not clear. In particular, the diagrams, which correspond to
 vector meson production (which dominates the higher twist small $x$
contribution in the Gribov theory, see discussion in Sect.~\ref{sec:ht}) seem
to be neglected as a very high twist effect.

\acknowledgments

This work was supported by the German-Israel Foundation (GIF), 
Sofia Kovalevskaya Program of the Alexander
von Humboldt Foundation (Germany) and the Department of Energy (USA).
 V.G. is grateful to Ingo Bojak for his help and explanations of QCDNUM.
 We also  thank H. Abramowicz for the discussion of the recent ZEUS
 diffractive data,
 and J. Morfin for emphasizing the need for a more reliable procedure for
 the extraction of
 $F_{2n}(x,Q^2)$ at small $x$.

\appendix
\section{Nuclear density $\rho_A$}
\label{sec:appendix}

 The nuclear density $\rho_A$, which enters the calculation of nuclear shadowing in 
Eq.~(\ref{eq:master2}), was parametrized in a two-parameter Fermi 
form for $^{40}$Ca, $^{110}$Pd, $^{197}$Au
and  $^{206}$Pb \cite{Vries}
\begin{equation}
\rho_{A}(r)=\frac{\rho_{0}}{1+\exp\left[(r-c)/a\right]} \ , 
\label{fermi}
\end{equation}
where $r=\sqrt{|\vec{b}|^2 +z^2}$ and
$a=0.545$ fm and the parameters $\rho_{0}$ and $c$ are presented in Table~\ref{table:fermi}.
 Also note that $\rho_{A}(\vec{b},z)$ was normalized as
 $2 \pi \int^{\infty}_{0} d|\vec{b}| \int^{\infty}_{-\infty} dz |\vec{b}| \rho_{A}(\vec{b},z)=1$.

\begin{table}
\begin{tabular}{|c|c|c|}
\hline
Nucleus & $\rho_{0}$ (fm$^{-3})$ & $c$ (fm) \\ \hline
 $^{40}$Ca & 0.0039769 & 3.6663\\
 $^{110}$Pd & 0.0014458 & 5.308\\
$^{197}$Au & 0.000808 & 6.516\\
 $^{206}$Pb & 0.0007720 & 6.6178\\ \hline
\end{tabular}
\caption{The parameters entering the nuclear one-body density $\rho_A (r)$.}
\label{table:fermi}
\end{table}

For $^{12}$C, we used
\begin{equation}
\rho_{A}(r)=\rho_{0}\left(1+\alpha \left(\frac{r}{a}\right)^2 \right)e^{-r^2/a^2} \, 
\label{mho}
\end{equation}
with $\rho_{0}=0.0132$, $\alpha=1.403$ and $a=1.635$ fm.

\section{Subleading (Reggeon) contribution to nuclear shadowing}
\label{sec:reggeon}

The analysis of the 1994 H1 data on hard diffraction was carried out with
the assumption that the diffractive structure function $F_2^{D(3)}$ is 
described by a sum of the effective Pomeron (leading) and
 Reggeon (subleading) contributions \cite{H1:1994}
\begin{equation}
F_2^{D(3)}(x_{\Pomeron}, \beta,Q^2)=f_{\Pomeron/p}(x_{\Pomeron}) \, F_2^{\Pomeron}(\beta,Q^2)+f_{\Reggeon/p}(x_{\Pomeron}) \, F_2^{\Reggeon}(\beta,Q^2) \,,
\label{regge1}
\end{equation}
where $f_{\Pomeron/p}$ and $f_{\Reggeon/p}$ are the so-called Pomeron and Reggeon fluxes; $F_2^{\Pomeron}$ and $F_2^{\Reggeon}$ are the Pomeron and Reggeon 
structure functions.
It is important to note that the both terms in Eq.~(\ref{regge1}) 
are leading twist contributions.
In our analysis we use fit 3 of model B of 
\cite{H1:1994} which assumed no interference between the leading and subleading
contributions.
 The latter contribution
 becomes important only for large values of measured 
$x_{\Pomeron}$, $x_{\Pomeron} > 0.01$.

In the H1 QCD analysis, $F_2^{\Reggeon}$ was assumed to be  the pion 
$F_2^{\pi}$
\cite{pion} multiplied by a free coefficient $C_{\Reggeon}$ to be determined
from the data. Unfortunately, the value of $C_{\Reggeon}$ is not given in 
the H1 publication. 
Therefore, using Eq.~(\ref{regge1}) we performed a $\chi^2$ fit to a set of
 selected H1 data points with $x_{\Pomeron} > 0.01$ and found that
$C_{\Reggeon} \approx 17$.

The Reggeon contribution to the nuclear shadowing correction to the $F_2^A$
 structure function has the form similar to our master 
equation~(\ref{eq:master2})
\begin{eqnarray}
&&\delta F_2^{A \, (\Reggeon)}(x,Q^2)=\frac{A(A-1)}{2} 16 \pi C_{\Reggeon}  {\cal R}e \Bigg[\frac{(1-i\eta_{\Reggeon})^2}{1+\eta_{\Reggeon}^2} 
\int d^2 b \int^{\infty}_{-\infty} dz_1 \int^{\infty}_{z_1} dz_2 \int^{0.1}_{x} d x_{\Pomeron}  
\nonumber\\
&&\times F_2^{\pi}(\beta,Q^2) \rho_A(b,z_1) \rho_A(b,z_2) e^{i x_{\Pomeron} m_N (z_1-z_2)} e^{-(A/2)(1-i\eta_{\Reggeon})\sigma_{{\rm eff}}^{\Reggeon} \int_{z_1}^{z_2} dz \rho_A(b,z)} \Bigg] \,.
\label{regge2} 
\end{eqnarray}
In this equation, the $\sigma_{{\rm eff}}^{\Reggeon}$ rescattering cross 
section is defined as (compare to Eq.~(\ref{eq:sigma}))
\begin{equation}
\sigma_{{\rm eff}}^{\Reggeon}(x,Q^2)=\frac{16 \pi C_{\Reggeon}}{F_2^N(x,Q^2)(1+\eta_{\Reggeon}^2)} \int_x^{0.1}d x_{\Pomeron}  F_2^{\pi}(\beta,Q^2) \,,
\label{regge3} 
\end{equation}
where it is worth noting the absence of the Reggeon flux since
$f_{\Reggeon/p}(x_{\Pomeron},t=0)=1$ \cite{H1:1994}.

The theoretical uncertainty associated with the Reggeon contribution
originates from the uncertainty in the choice of $\eta_{\Reggeon}$, which is
the ratio of the real to imaginary parts of the subleading exchange 
amplitude,
\begin{equation}
\eta_{\Reggeon}=-\frac{\xi+\cos \pi  \alpha_{\Reggeon}(0)}{\sin \pi \alpha_{\Reggeon}(0)}\,,
\end{equation}
where $\xi=\pm 1$ is the signature factor. The intercept of the Reggeon
trajectory, $\alpha_{\Reggeon}(0)$, was a fit parameter in the H1 analysis.
The fit to the H1 diffractive data gives
$\alpha_{\Reggeon}(0)=0.5 \pm 0.11 (stat.) \pm 0.11(sys.)$ \cite{H1:1994}, 
which leads to $\eta_{\Reggeon}=\pm 1$ depending on the signature factor.
For the $\rho$ and $\omega$ meson exchanges, $\eta_{\Reggeon}=1$;
for the $a$ and $f$ meson exchanges, $\eta_{\Reggeon}=-1$.
Since no attempt was made to separate the contributions with different
signatures to the Reggeon contribution in the analysis of \cite{H1:1994} 
(inclusive diffraction is not sensitive to the signature of the exchange)
 the value of  $\eta_{\Reggeon}$ is uncertain, 
$-1 \leq \eta_{\Reggeon} \leq 1$.

\begin{figure}
\includegraphics[width=13cm,height=13cm]{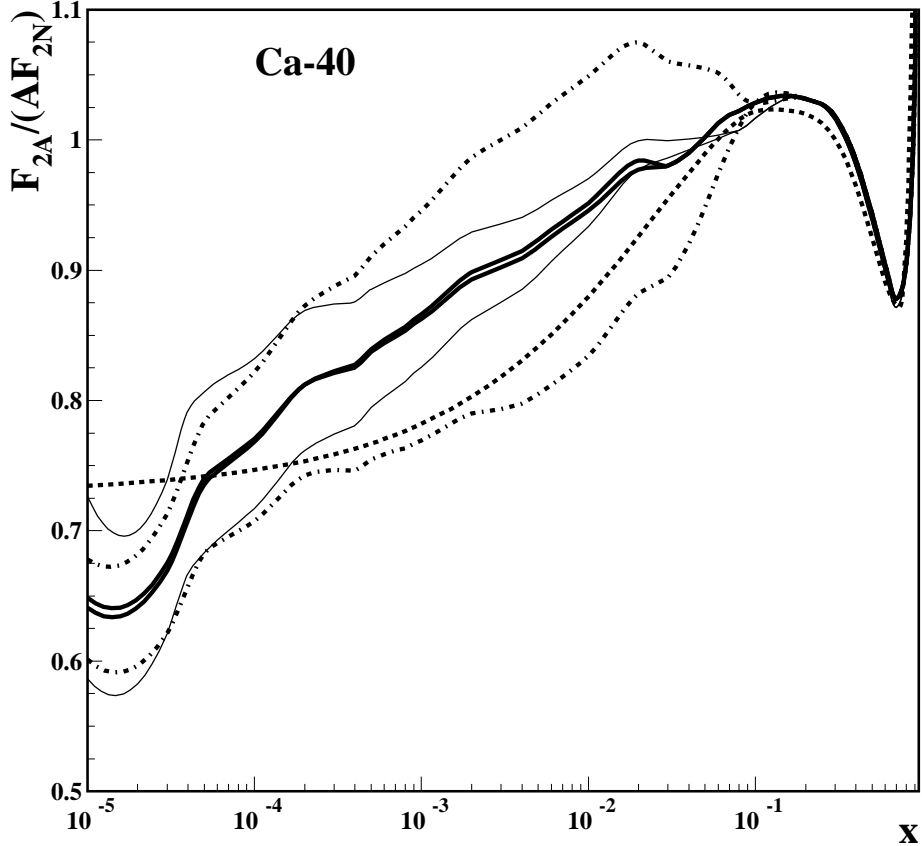}
\caption{The $F_2^A/(A F_2^N)$ ratio at $Q^2=4$ GeV$^2$ for Ca-40. 
The thick solid curves present our main result; 
the thin solid curves present the uncertainty of our predictions;
the dashed curve presents the fit of \cite{Eskola}; the dot-dashed curves 
is the result of the calculation including both the Pomeron and Reggeon
contributions to diffraction. The lower dot-dashed curve corresponds to
$\eta_{\Reggeon}=-1$; the upper one corresponds to $\eta_{\Reggeon}=1$.
}
\label{fig:reggeon_a40}
\end{figure}

Because of the large $|\eta_{\Reggeon}|$, the uncertainty in the choice of
$\eta_{\Reggeon}$ leads to a very significant uncertainty in the resulting
shadowing correction. This is illustrated in Fig.~\ref{fig:reggeon_a40}
presenting $F_2^A/(AF_2^N)$ for Ca-40 at $Q^2=4$ GeV$^2$.
In Fig.~\ref{fig:reggeon_a40}, the thick solid curves present the predictions
of our model neglecting the subleading exchange contribution to hard diffraction; the thin solid curves present the uncertainty of the predictions;
the dashed curve presents the result of \cite{Eskola}; the dot-dashed curves 
is the result of the calculation including both the Pomeron and Reggeon
contributions to diffraction. The lower dot-dashed curve corresponds to
$\eta_{\Reggeon}=-1$; the upper one corresponds to $\eta_{\Reggeon}=1$.
As seen from Fig.~\ref{fig:reggeon_a40}, the variation of $\eta_{\Reggeon}$
between its lower and upper limits leads to a dramatic change in the
predicted nuclear shadowing.

\begin{figure}
\includegraphics[width=13cm,height=13cm]{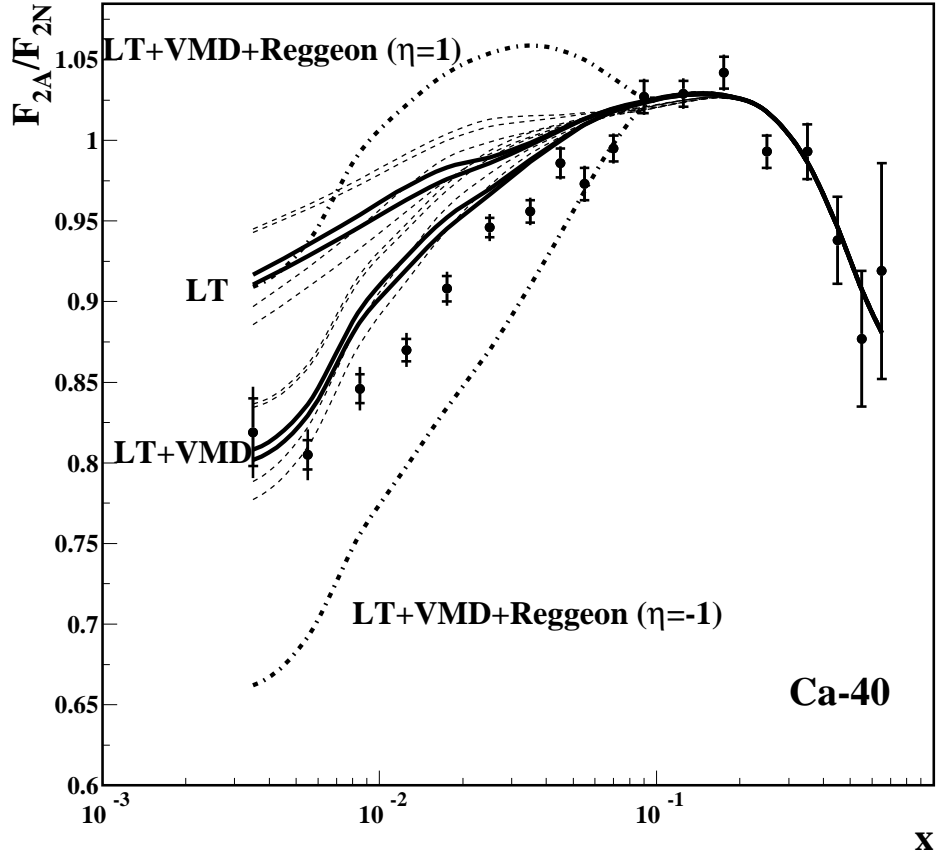}
\caption{Comparison of the leading twist theory results (the upper set 
of solid curves and associated dashed error bands)
to the NMC data on $F_2^{Ca}/F_2^N$ \protect\cite{NMC1}.
The lower set of the solid curves is obtained by adding the VMD contribution
using Eq.~(\ref{eq:vmd}).
The dot-dashed curves are the sums of the Pomeron, Reggeon and VMD contributions. The lower dot-dashed curve corresponds to
$\eta_{\Reggeon}=-1$; the upper one corresponds to $\eta_{\Reggeon}=1$.
}
\label{fig:reggeon_catod}
\end{figure}

\begin{figure}
\includegraphics[width=13cm,height=13cm]{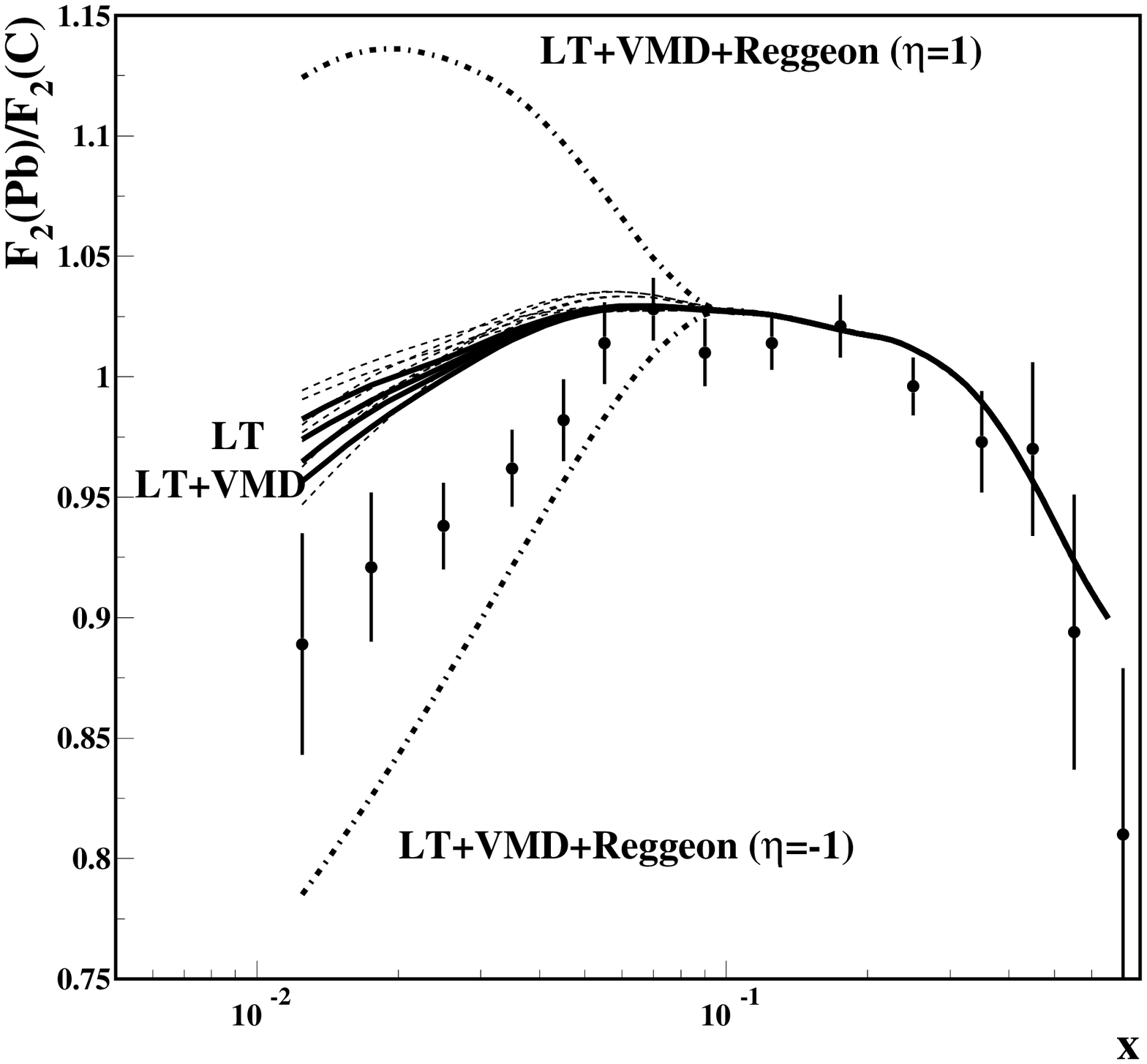}
\caption{Comparison of the leading twist theory results (the upper set 
of solid curves and associated dashed error bands)
to the NMC data on $F_2^{Pb}/F_2^C$ \protect\cite{NMC2}.
The lower set of the solid curves is obtained by adding the VMD contribution
using Eq.~(\ref{eq:vmd}).
The dot-dashed curves are the sums of the Pomeron, Reggeon and VMD contributions. The lower dot-dashed curve corresponds to
$\eta_{\Reggeon}=-1$; the upper one corresponds to $\eta_{\Reggeon}=1$.
}
\label{fig:reggeon_pbtoc}
\end{figure}

The influence of the Reggeon contribution on the comparison of our predictions
to the NMC fixed-target nuclear DIS data is presented in Figs.~\ref{fig:reggeon_catod} and \ref{fig:reggeon_pbtoc}.
The upper set of solid curves and the associated dashed error bands
is the result of the calculation using only the Pomeron contribution 
to diffraction.
The lower set of the solid curves is obtained by adding the VMD contribution
using Eq.~(\ref{eq:vmd}).
The dot-dashed curves are the results of the calculation taking
into account the Pomeron, Reggeon and VMD contributions. 
The lower dot-dashed curve corresponds to
$\eta_{\Reggeon}=-1$; the upper one corresponds to $\eta_{\Reggeon}=1$.

As seen from Figs.~\ref{fig:reggeon_catod} and \ref{fig:reggeon_pbtoc},
varying $\eta_{\Reggeon}$ in the $-1 \leq \eta_{\Reggeon} \leq 1$ range,
one obtains a wide spectrum of predictions for $F_2^{Ca}/F_2^D$ 
and $F_2^{Pb}/F_2^C$, which accommodate the NMC data. This indicates the 
possibility of a duality between the higher twist vector meson contribution
and the leading twist subleading contribution.

\end{document}